\documentclass[lettersize,journal]{IEEEtran}

\usepackage{amsmath,amsfonts,amssymb}
\usepackage{algorithm}
\usepackage{algpseudocode}
\usepackage{array}
\usepackage[caption=false,font=normalsize,labelfont=sf,textfont=sf]{subfig}
\usepackage{textcomp}
\usepackage{stfloats}
\usepackage{url}
\usepackage{cite}
\usepackage{graphicx}
\usepackage{mathtools}
\usepackage{booktabs}
\usepackage{pifont}
\usepackage{multirow}
\usepackage{tikz}
\usepackage{enumitem}
\usepackage{hyperref}
\usepackage{float} 
\usepackage{comment}
\usepackage{amsthm}

\usepackage{tcolorbox} 
\usepackage[table]{xcolor}

\tcbuselibrary{skins}
\tcbset{
  defbox/.style={
    enhanced, frame hidden, sharp corners,
    colback=black!3,
    borderline west={1pt}{0pt}{black!55},
    left=8pt, right=6pt, top=4pt, bottom=4pt,
    before skip=6pt, after skip=6pt
  }
}

\newcommand{\cmark}{\ding{51}}
\newcommand{\xmark}{\ding{55}}
\definecolor{shragrow}{RGB}{235,240,250}
\definecolor{hdrblue}{RGB}{220,230,245}
\begin{document}

\title{SHACR: A Graph-Augmented Semi-Autonomous Framework for Multi-Class
       Conflict Resolution in Smart Home IoT Automation}

\author{%
  Leena~Marghalani,~\IEEEmembership{}%
  Walid~Aljoby,~\IEEEmembership{Member,~IEEE, }%
  Suayb~S.~Arslan,~\IEEEmembership{Senior Member,~IEEE}%

  \thanks{L.~Marghalani and W.~Aljoby are with the Department of Information and Computer Science, King Fahd University
    of Petroleum and Minerals (KFUPM), Dhahran 31261, Saudi Arabia
    (e-mail: g202303130@kfupm.edu.sa; waleed.algobi@kfupm.edu.sa;).}
  \thanks{S.~S.~Arslan is a Professor in the Department of Computer
    Engineering and Director of the Institute for Data Science and AI,
    Bo\u{g}azi\c{c}i University, Istanbul 34342, Turkey, and a Research Affiliate in the Department of Brain and Cognitive Sciences,
    Massachusetts Institute of Technology, Cambridge, MA, USA
    (e-mail: suayb.arslan@bogazici.edu.tr, sarslan@mit.edu).}
  \thanks{Manuscript received XX Month XXXX; revised XX Month XXXX;
    accepted XX Month XXXX.
    This work was supported in part by [funding information].
    \emph{(Corresponding author: Walid~Aljoby.)}}}

\markboth{IEEE Internet of Things Journal}%
{Marghalani \MakeLowercase{\textit{et~al.}}: SHACR---Semi-Autonomous Framework for
Smart Home IoT Conflict Resolution}

\maketitle

\begin{abstract}
Smart home automation increasingly relies on user-defined rules across heterogeneous 
IoT devices. While these rules appear harmless in isolation, their concurrent execution 
creates hidden, cross-rule interactions via shared devices, environmental variables, 
and physical topology. These interactions result in unsafe, wasteful, or 
privacy-threatening behaviors that are completely invisible to text-only analysis. 
Existing conflict detectors remain siloed, catching either static syntactic conflicts or specific environment-mediated interactions without unifying the two or providing actionable repairs for non-expert users.
We present \textsc{SHACR}, a smart home conflict resolution framework that anchors Large Language Model (LLM) unpredictability by grounding its reasoning in a formal, 
directed knowledge graph. \textsc{SHACR} encodes devices, capabilities, physical states, 
and Trigger--Condition--Action rules as typed, traversable entities. By elevating 
physical cause-effect relationships to first-class graph edges, \textsc{SHACR} transforms 
conflict detection from fragile text inference into deterministic multi-hop graph traversal, 
unifying logical, semantic, and physical conflict classes. It drives a closed-loop Scan-Explain-Repair-Validate workflow that uses the graph to bound the LLM's action space. We evaluated \textsc{SHACR} on a testbed of 203 rules deployed across 70 apartments within a smart building. By holding the underlying LLM fixed and introducing \textsc{SHACR}'s knowledge graph, classification errors drop by 36.7\%, F1 rises from 0.59 to 0.79, and few-shot calibration further lifts F1 to 0.95, whereas the same calibration barely helps a graph-free LLM. Ultimately, this work challenges the current AI paradigm, establishing that structured knowledge representation is a far more critical factor for dependable IoT automation management than prompt engineering or underlying model architecture.
\end{abstract}

\begin{IEEEkeywords}
Smart homes, Internet of Things, knowledge graphs, graph retrieval-augmented
generation, large language models, trigger-condition-action rules, conflict
detection, conflict resolution, IoT automation, autonomous systems, agentic RAG.
\end{IEEEkeywords}

\section{Introduction}
\label{sec:intro}

\IEEEPARstart{S}{mart} homes have become one of the most pervasive deployments
of AI-enabled IoT, with residents routinely combining sensors, appliances, and
cloud services to automate lighting, climate control, access control, and
security monitoring~\cite{ding2021iotsafe}. Unlike enterprise IoT environments,
smart homes are configured and maintained by end users---not trained
administrators---often across multiple applications and vendors~\cite{philip2023theres,brush2011home,Becks2023ComplexitySmartHome,He2019SmartDevicesStupid,Hao2024FrustrationToFunction}.
As automation rule sets grow, routines that appear benign in isolation can
interact in unsafe, wasteful, or privacy-threatening ways once they share rooms,
devices, environmental variables, or cross-platform triggers~\cite{ding2021iotsafe,
chen2024tapchecker,chen2019multi,chi2023detecting}.

Consider a concrete failure scenario. An energy-saving rule turns off a living-room smart
plug at night. A surveillance rule activates a security camera when the home
enters \emph{away} mode. If the camera draws power through that plug, the first
rule silently disables the second. No rule text states \emph{``turn off the
camera,''} yet the home loses surveillance precisely when monitoring matters
most. This failure is invisible to any analysis confined to rule syntax, because
the dependency lives in the physical wiring, not in the logic of either rule.

The same failure mechanism recurs across many everyday configurations. A robot
vacuum passing through the entrance hallway triggers the same motion sensor used
to detect human presence, satisfying a precondition that unlocks the front door
while the apartment is unoccupied. A heater rule raises room temperature past
the threshold that opens a motorized window, which then cools the room and
reactivates the heater---an energy-wasting feedback loop that neither rule
individually encodes. A scheduled oven rule forces the appliance on while a
temperature-safety rule attempts to shut it off, producing a thermal overload
hazard through conflicting actuation. In each case, the conflict is not a
programming error within any single rule; it is an emergent interaction across
rules that were each authored independently and each appear reasonable alone.

The fundamental question this work addresses is whether, given a set of heterogeneous smart home automation rules, device placements, and environmental dependencies, it is possible to determine prior to deployment whether any subset of rules can jointly produce unsafe or unexpected behavior. It further asks whether such conflicts can be explained in terms that non-expert users can understand and act on. Existing approaches only
partially address this issue. Formal rule-centric analyzers~\cite{al2019iotc,chen2024tapchecker}
detect syntactic contradictions---two rules issuing incompatible commands to the
same actuator---but miss conflicts mediated by device topology or environmental
propagation. Semantic frameworks~\cite{wang2020iot,chi2023detecting,ding2021iotsafe}
model how device actions propagate through shared physical variables, but
typically require manually specified device behavior models and provide neither
cross-platform unification nor user-facing repair guidance. Recent LLM-based
approaches~\cite{giudici2025generating,rey2024leveraging,jahanbakhsh2025leveraging}
improve natural-language interaction but remain unreliable when critical topology
and dependency information is absent from the prompt context.

Conflicts arise because reasoning, confined to rule text, is structurally blind to
three qualitatively different failure classes (\S\ref{sec:bg}), and these failures
can be avoided if rules are reasoned over a structured representation that makes
physical dependencies explicit and traversable (\S\ref{sec:model}). We therefore
present \textsc{SHACR} (Smart Home Semi-Autonomous Conflict Resolution), a
semi-autonomous framework built on \emph{graph-grounded conflict detection}: the
principle that detecting, explaining, and repairing automation conflicts requires
anchoring LLM reasoning in a formal knowledge graph that encodes device placement,
capability effects, environmental variables, and Trigger--Condition--Action (TCA)
rule structure as typed, directed entities. By making physical cause-effect
relationships first-class graph edges, \textsc{SHACR} enables detection of all
three conflict classes through structured multi-hop traversal rather than text
inference, and enables explanation by tracing the traversal path back to the
specific devices and rules that caused the failure.

\textsc{SHACR} achieves this through a tightly integrated architecture that goes
well beyond GraphRAG retrieval alone. It comprises a structured data ingestion
pipeline that normalizes heterogeneous YAML rule specifications into the knowledge
graph, a Neo4j-hosted knowledge graph store, a two-tier MCP server layer that
exposes five purpose-built domain tools to the reasoning engine, a
Scan--Explain--Repair--Validate workflow that orchestrates multi-step LLM
inference over retrieved subgraphs, and a Streamlit dashboard that enforces
human approval before any graph modification executes. Each component addresses
a specific failure mode in the broader conflict management lifecycle.

We prototype \textsc{SHACR} on Neo4j AuraDB with a custom MCP server layer and
a Streamlit user dashboard and evaluate it on a testbed consisting 
of 203 automation rules across 70 apartments, demonstrating an F1-score of 
0.79 in a zero-shot configuration. This represents a 36.7\% reduction in 
classification errors over an LLM-only baseline using the same model, which 
increases to 0.95~F1 with balanced few-shot prompting.

The principal contributions of this work are as follows.
\begin{enumerate}[leftmargin=*,label=\arabic*.]
  \item We characterize three qualitatively distinct smart home conflict
  classes (logical, semantic, and physical) and show that each arises from
  a structurally different failure mechanism in IoT automation. We further
  demonstrate that existing approaches detect at most one of these classes,
  establishing the need for a unified, graph-grounded detection framework
  (\S\ref{sec:bg}).
  \item We formalize the three conflict classes over a directed knowledge graph
  $G$ and show that each maps to a specific traversal pattern over the
  \texttt{AFFECTS} edge---the physical cause-effect substrate that makes
  device-mediated interactions graph-detectable before deployment
  (\S\ref{sec:model}).
  \item We design and prototype \textsc{SHACR}, a semi-autonomous conflict
  management system comprising a structured data ingestion pipeline, a typed
  knowledge graph schema, a GraphRAG pipeline with a five-tool MCP server layer,
  and a Scan--Explain--Repair--Validate workflow that proposes structured repairs
  within a constrained action space and verifies resolution against the live
  graph (\S\ref{sec:design}--\S\ref{sec:impl}).
  \item We evaluate \textsc{SHACR} on a controlled testbed of 203 automation
  rules across 70 apartments and show a 36.7\% reduction in total classification
  errors over an LLM-only baseline using the same underlying model, an F1-score
  of 0.95 with balanced few-shot prompting, and present a cross-model comparison of Claude Sonnet~4 and Google Gemini under matched prompting conditions. To support reproducibility, the full 203-rule, 70-apartment testbed and ingestion code are released publicly (\S\ref{sec:eval}).
\end{enumerate}

The paper is organized as follows.
Section~\ref{sec:related} surveys related work.
Section~\ref{sec:bg} presents background and motivation.
Section~\ref{sec:model} formalizes the conflict taxonomy.
Section~\ref{sec:design} describes the system design.
Section~\ref{sec:impl} covers the prototype implementation.
Section~\ref{sec:eval} presents the evaluation.
Section~\ref{sec:disc} discusses findings and future direction.
Section~\ref{sec:conclusion} concludes.

\section{Related Work}
\label{sec:related}

\textbf{Logical conflict detection.}
IoTC$^2$~\cite{al2019iotc} applies formal methods to identify rule violations
against predefined safety properties. TapChecker~\cite{chen2024tapchecker} uses
SMT-based analysis for trigger-action conflicts. Fine-Grained Conflict
Detection~\cite{chaki2020fine} and IoTSAFE~\cite{ding2021iotsafe} extend this
with service-level and policy-level formulations. These methods reliably surface
direct contradictions but are structurally limited to syntactic rule-level
analysis; conflicts mediated by device topology or environmental propagation
remain invisible.

\textbf{Semantic and hybrid detection.}
The IoTMon~\cite{ding2018safety} introduces physical interaction channels.
The IoT-Praetor~\cite{wang2020iot} models the environmental influence. SeIoT~\cite{li2024seiot} uses knowledge graphs for semantic anomaly detection, and contextual semantics of behavior-graphs have also been applied to anomaly detection in smart IoT systems~\cite{lin2024anomaly}. The IoTMediator~\cite{chi2023detecting}
and SAFE-TAP~\cite{kuang2025safe} address cross-platform threats and semantic
embeddings, respectively. VISCR~\cite{nagendra2019viscr} and IoTIE~\cite{chen2019multi}
combine rule-based and contextual reasoning for cross-platform detection.
AutoIoT~\cite{cheng2024autoiot} couples LLMs with formal verification for rule
generation. Although these systems demonstrate that IoT devices interact through
both cyber and physical channels, they rely on predefined behavior models, lack
automated repair mechanisms, and provide limited user-facing explanation.

\textbf{LLM-based and retrieval-augmented approaches.}
ChatIoT~\cite{dong2025chatiot} integrates LLM reasoning with threat
intelligence. Rey et al.~\cite{rey2024leveraging} and Jahanbakhsh et
al.~\cite{jahanbakhsh2025leveraging} use retrieval for personalization and
workflow composition. LLM-HA~\cite{giudici2025generating} provides explanation
for HomeAssistant automations. These systems improve contextual reasoning but
rely on textual descriptions and cannot reliably detect multi-hop or
physically-mediated conflicts without an explicit relational model.

\textbf{GraphRAG and agentic RAG.}
\textsc{SHACR} is a knowledge-based GraphRAG system~\cite{zhang2025survey} in
which the knowledge graph is the primary carrier of domain semantics, enabling
evidence aggregation along explicit relational paths rather than surface-level
lexical similarity. Beyond static GraphRAG, \textsc{SHACR} corresponds to the
\emph{corrective single-agent} agentic RAG architecture~\cite{singh2025agentic}:
a centralized reasoning agent iteratively refines its outputs through a
structured repair-and-validate loop, grounded in a formal typed graph rather
than unstructured document corpora, and gated by explicit user approval before
any knowledge graph modification executes.

\textbf{Positioning.}
\textsc{SHACR} differs from all prior work in encoding physical cause-effect
relationships as first-class, traversable graph edges, and in combining
structured retrieval, multi-class conflict detection, executable repair, and a
validated semi-autonomous pipeline within a single integrated system.
Table~\ref{tab:related_work_comparison} maps capabilities across representative
frameworks. No prior system simultaneously handles all three conflict classes
across heterogeneous platforms while providing automated mitigation, grounded
explanation, and LLM-based reasoning within a user-approved repair loop; all
existing systems reason over rule text or predefined device models rather than a
persistent typed knowledge graph that makes physical dependencies traversable.

\begin{table}[t]
  \centering
  \caption{Capability comparison across representative smart home conflict
  detection frameworks.}
  \label{tab:related_work_comparison}
  \setlength{\tabcolsep}{2.5pt}
  \renewcommand{\arraystretch}{1.0}
  \footnotesize
  \resizebox{\columnwidth}{!}{%
  \begin{tabular}{@{}l c c c c c c c c c@{}}
    \toprule
    \multirow{2}{*}{\textbf{System}} &
    \multirow{2}{*}{\textbf{Year}} &
    \multicolumn{4}{c}{\textbf{Conflict Analysis}} &
    \multicolumn{4}{c}{\textbf{Capabilities}} \\
    \cmidrule(lr){3-6}\cmidrule(lr){7-10}
    & &
    \shortstack{Logical} & \shortstack{Semantic} &
    \shortstack{Env.\\Effects} & \shortstack{Multi-\\Plat.} &
    Mitigation & \shortstack{Explain-\\able} & \shortstack{LLM-\\based} &
    \textcolor{black}{\shortstack{Graph-\\Grounded}} \\
    \midrule
    IoTC$^2$~\cite{al2019iotc}           & 2019 & \cmark & \xmark & \xmark & \xmark & \xmark & \xmark & \xmark & \textcolor{black}{\xmark} \\
    VISCR~\cite{nagendra2019viscr}       & 2019 & \cmark & \cmark & \xmark & \cmark & \cmark & \xmark & \xmark & \textcolor{black}{\xmark} \\
    IoTIE~\cite{chen2019multi}           & 2019 & \cmark & \cmark & \xmark & \cmark & \xmark & \xmark & \xmark & \textcolor{black}{\xmark} \\
    IoT-Praetor~\cite{wang2020iot}       & 2020 & \xmark & \cmark & \cmark & \xmark & \xmark & \xmark & \xmark & \textcolor{black}{\xmark} \\
    IoTSAFE~\cite{ding2021iotsafe}       & 2021 & \xmark & \cmark & \cmark & \xmark & \xmark & \cmark & \xmark & \textcolor{black}{\xmark} \\
    IoTMediator~\cite{chi2023detecting}  & 2023 & \xmark & \cmark & \cmark & \cmark & \xmark & \cmark & \xmark & \textcolor{black}{\xmark} \\
    AutoIoT~\cite{cheng2024autoiot}      & 2024 & \cmark & \cmark & \xmark & \cmark & \cmark & \xmark & \cmark & \textcolor{black}{\xmark} \\
    TapChecker~\cite{chen2024tapchecker} & 2024 & \cmark & \xmark & \xmark & \xmark & \xmark & \xmark & \xmark & \textcolor{black}{\xmark} \\
    SAFE-TAP~\cite{kuang2025safe}        & 2025 & \cmark & \cmark & \cmark & \xmark & \xmark & \xmark & \cmark & \textcolor{black}{\xmark} \\
    LLM-HA~\cite{giudici2025generating}  & 2025 & \xmark & \xmark & \xmark & \xmark & \xmark & \cmark & \cmark & \textcolor{black}{\xmark} \\
    \midrule
    \rowcolor{shragrow}
    \textbf{SHACR (ours)} & \textbf{2026} &
    \cmark & \cmark & \cmark & \cmark & \cmark & \cmark & \cmark & \textcolor{black}{\cmark} \\
    \bottomrule
  \end{tabular}}
\end{table}

\section{Background and Motivation}
\label{sec:bg}

\subsection{What Smart Home Automation Rules Do}
\label{subsec:what_rules_do}

A smart home automation system can be understood as a collection of reactive
control policies, each expressed as a TCA rule.
The trigger specifies an observable event or state transition---motion
detected, temperature falling below a threshold, time reaching a scheduled
point---that causes the rule to become eligible for evaluation. The condition
constrains when the rule should actually fire, filtering eligibility by additional
predicates such as the home mode (\emph{Away}, \emph{Night}) or a device's
current state. The action specifies the command issued once the rule fires:
turning a device on or off, adjusting a set-point, sending a notification,
or locking a physical actuator.

Individual rules are typically authored for specific, narrow purposes:
an energy-saving schedule turns off idle devices at night; an occupancy
rule enables lights when motion is detected; a safety rule locks the front
door when no presence is reported~\cite{hazazi2024exploring}. Considered in isolation, each of these
policies is self-consistent and straightforwardly correct. The difficulty
arises when multiple rules, each authored independently by a non-expert
user, begin to share devices, rooms, environmental variables, or derived
contextual predicates. At that point, the interaction space grows in ways
that are not visible from any individual rule's text.

To make this concrete, consider the scenario illustrated in
Figure~\ref{fig:subgraph_conflict_example}. A smart apartment entrance has
three rules in deployment. The first, \emph{Robot Vacuum Daily Run}, schedules
the vacuum to clean the apartment on a recurring basis. The second, \emph{Entrance
Presence}, sets a virtual presence flag whenever the entrance motion sensor
detects movement. The third, \emph{Unlock on Presence}, unlocks the front door
whenever that presence flag becomes active during daytime hours. Reading each
rule individually, a resident sees three sensible automations: a cleaning
schedule, a presence monitor, and a door-unlock convenience. Nothing in any
single rule's text indicates a problem.

The conflict, however, is real. During its scheduled cleaning cycle, the
vacuum passes through the entrance hallway. As it does, it triggers the
entrance motion sensor---the same sensor that the presence rule monitors.
The presence rule activates the virtual flag. The unlock rule finds its
daytime condition satisfied and issues an unlock command to the front door.
The apartment is now unsecured, and no single rule bears the immediate
cause: the failure exists only in the \emph{composition} of the three rules
across the shared sensor.

\subsection{How the Knowledge Graph Makes Conflicts Visible}
\label{subsec:kg_visibility}

This failure trace illustrates why representation matters. A text-only analysis
of the three rules sees three independent logical constructs that share no
direct syntactic dependency. The vacuum rule targets the vacuum's run
capability; the presence rule targets the motion sensor state; the unlock
rule targets the lock capability conditioned on the presence flag. Without
an explicit model of the physical world, no analysis can determine that the
vacuum's movement activates the motion sensor, which corrupts the semantic
precondition of the unlock rule.

\begin{figure}[t]
  \centering
  \includegraphics[width=0.45\textwidth]{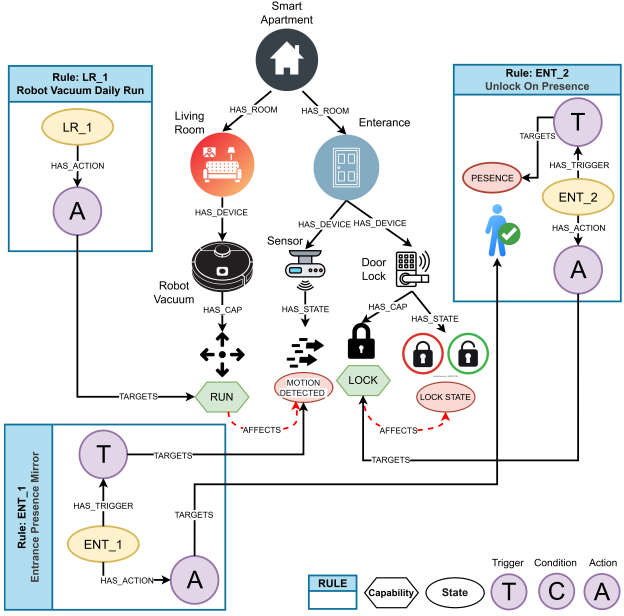}
  \caption{Knowledge graph subgraph of the running example. Three rules---Robot
  Vacuum Daily Run, Entrance Presence, and Unlock on Presence---appear
  independent in rule text but form a traceable causal chain when embedded as
  graph entities. The \texttt{AFFECTS} edge from the vacuum's \texttt{RUN}
  capability to \texttt{MOTION\_DETECTED} is the dependency absent from all
  rule texts yet sufficient to produce an unintended door unlock. Encoding this
  edge makes pre-deployment detection possible; tracing it back produces a
  grounded explanation the user can act on.}
  \label{fig:subgraph_conflict_example}
\end{figure}

Figure~\ref{fig:subgraph_conflict_example} shows how this failure becomes
tractable once the rules are embedded in a knowledge graph. The
\texttt{AFFECTS} edge from the vacuum's \texttt{RUN} capability to the
\texttt{MOTION\_DETECTED} state node encodes the physical fact that the
vacuum's movement produces sensor activations. This single edge---absent
from any rule's text but present in any accurate model of the deployed
environment---closes the causal chain: it makes the path from
\emph{vacuum running} $\to$ \emph{motion sensor fires} $\to$
\emph{presence flag set} $\to$ \emph{front door unlocked} a traversable
graph structure. Once that structure is explicit, the conflict is
graph-detectable before deployment.

The knowledge graph serves three complementary roles in \textsc{SHACR}.
As a \emph{dependency map}, the \texttt{AFFECTS} edge explicitly encodes
physical cause-effect relationships among device capabilities and environmental
or state variables that are otherwise invisible to text-based analysis. As a
\emph{retrieval index}, apartment-level subgraph extraction gives the LLM a
focused, contextually complete view of the relevant environment rather than a
flat list of rule texts. As an \emph{explanation structure}, detected conflicts
correspond to traversal paths in the graph, so the system can report the specific
nodes and edges responsible for the failure, producing explanations that users can
verify against their own home configuration and use to decide on a repair.

\subsection{How Conflicts Arise: Three Failure Mechanisms}
\label{subsec:how_conflicts}

The vacuum-door scenario is an instance of a broader taxonomy. Conflicts in
smart home automation arise through three structurally distinct mechanisms,
each requiring different reasoning to detect.

\textbf{Direct contradictions (logical conflicts).} Two rules issue mutually
exclusive commands to the same actuator under overlapping conditions. An
air-conditioning cooling rule and an energy-saving shutdown rule that both
target the same AC unit are a canonical example. These conflicts are detectable
by syntactic rule comparison alone: no model of physical propagation is
required because both rules reference the same device state node in their action
clauses.

\textbf{Environment-mediated conflicts (semantic conflicts).} One rule alters a
physical variable---temperature, illuminance, airflow, or smoke concentration---that
triggers a second rule, forming an implicit causal chain not encoded in either
rule's text. A heater rule that raises room temperature above the threshold that
opens a motorized window, which then cools the room and reactivates the heater,
creates an energy-wasting feedback loop that is invisible to syntactic analysis.
Neither rule is incorrect; the conflict emerges through the shared environmental
variable that mediates between them.

\textbf{Topology-mediated conflicts (physical conflicts).} One rule's action
changes a device state that serves as a precondition for a different rule through
a multi-hop chain of physical dependencies spanning heterogeneous entity types.
The vacuum-door scenario is precisely this class, as is the camera-plug
scenario: an energy-saving rule disables a smart plug, which cuts power to a
security camera, which defeats a surveillance rule operating on a separate
platform. No direct syntactic dependency connects the energy-saving rule to the
surveillance rule; the connection exists only through the physical wiring between
the plug and the camera.

These three mechanisms define the conflict classes that \textsc{SHACR} is
designed to detect, explain, and repair, formalized in \S\ref{sec:model}.

\subsection{Why Existing Approaches Fall Short}
\label{subsec:existing_fall_short}

Each existing approach addresses part of the problem. Formal rule-centric
analyzers~\cite{al2019iotc,chen2024tapchecker} handle direct contradictions
precisely and deterministically, but are structurally limited to syntactic
rule-level analysis. They have no model of how a vacuum's movement can trigger
a motion sensor, or how a heater's action affects a room temperature variable
that feeds another rule's trigger.

Semantic frameworks~\cite{wang2020iot,ding2018safety,chi2023detecting} extend
analysis to model physical propagation, but rely on predefined device behavior
models and manually specified environmental influence functions. They do not
generalize to new device types or cross-platform rule engines without significant
manual effort, and they provide limited support for user-facing explanation or
automated repair. Recent LLM-based
approaches~\cite{giudici2025generating,rey2024leveraging,jahanbakhsh2025leveraging}
improve natural-language interaction but remain unreliable when critical topology
and dependency information is absent from the prompt: without an explicit model
of how devices connect through power, sensor signals, and shared environments,
a language model cannot reliably infer that a vacuum's cleaning route could
unlock the front door.

The limitation shared by all existing approaches is that they reason over rule
text. Even language models that can interpret natural language cannot reliably
infer physical device dependencies from rule descriptions alone, because those
dependencies exist in the deployment environment---not in the rule logic. A
structured representation is needed that makes three properties explicit:
(1)~\emph{where} devices are deployed in the home, (2)~\emph{what} physical
effects their capabilities have on states and environmental variables, and
(3)~\emph{how} automation rules reference those entities through their TCA
structure. With such a representation, detecting a conflict becomes a graph
traversal problem rather than an inference problem, and explaining it becomes a
matter of returning the traversal path to the user.

\section{Conflict Taxonomy and Formal Definitions}
\label{sec:model}

\subsection{Notation}
\label{subsec:notation}

The following notation is used uniformly throughout the formal definitions in
this section. The smart home is represented as a directed, typed knowledge graph
$G=(V,E,\tau_V,\tau_E,A_V,A_E)$, where $V$ is the finite set of nodes and
$E \subseteq V \times V$ is the set of directed edges. The function
$\tau_V : V \to \mathcal{T}_V$ assigns each node a type drawn from the node
type vocabulary $\mathcal{T}_V$ (e.g., \texttt{Rule}, \texttt{Capability},
\texttt{State}, \texttt{EVar}); analogously, $\tau_E : E \to \mathcal{T}_E$
assigns each edge a type drawn from the edge type vocabulary $\mathcal{T}_E$
(e.g., \texttt{AFFECTS}, \texttt{TARGETS}, \texttt{HAS\_ACTION}). The
attribute functions $A_V : V \to \mathcal{A}$ and $A_E : E \to \mathcal{A}$
map nodes and edges to their associated attribute dictionaries; attribute
access is written $A_V(v).\mathit{attr}$ for a specific attribute field.

In the definitions that follow, $r_i$ and $r_j$ denote distinct
\texttt{Rule} nodes; $a_i$ and $a_j$ denote their respective \texttt{Action}
sub-nodes; $s$ denotes a \texttt{State} node that serves as a shared actuation
target; $e$ denotes an \texttt{EVar} node representing a room-scoped
environmental variable; $c$ denotes a \texttt{Capability} node whose actuation
propagates effects downstream; and $t_j$ denotes the \texttt{Trigger} or
\texttt{Condition} sub-node of $r_j$ that observes the resulting state or
variable. The safety predicate $\Phi(G)$ is a system-specific logical formula
over the graph state (e.g., \emph{``the front door lock is engaged whenever no
authorized presence is detected''}); a physical conflict is defined as a
composition of effects along a multi-hop path that causes $\Phi(G)$ to be
violated.

\subsection{Knowledge Graph Schema}
\label{subsec:schema}

The schema is organized into three semantic layers, summarized in
Table~\ref{tab:node_types}. The layered design reflects the natural structure
of the smart home: a physical topology layer describes where things are and what
they are; a device capability layer describes what those things can do and what
they observe; and an automation layer describes how rules reference and act upon
all of the above. Each layer exists not only for representational completeness
but because the conflict classes defined below require reasoning that cuts across
all three layers simultaneously.

\emph{Topological layer.} A \texttt{Building} node is the root entity containing
one or more \texttt{Apartment} nodes. Each apartment is subdivided into
\texttt{Room} nodes representing the spatial units of the deployment (living
room, bedroom, kitchen, entrance, etc.). Rooms serve as retrieval anchors
for apartment-level subgraph extraction and are the primary unit of analysis
for conflict localization: a conflict reported in the context of a specific room
can be communicated to the user in terms they recognize from their daily
experience of the space.

\emph{Device layer.} A \texttt{Device} node represents a physical IoT sensor or
actuator installed in a room. Each device exposes one or more \texttt{Capability}
nodes that encode the abstract operations it can perform (e.g., \texttt{power},
\texttt{locking}, \texttt{heating}, \texttt{recording}, \texttt{cooling},
\texttt{dimming}). Capabilities are the primary targets of automation actions:
when a rule fires, it actuates a capability, not a device directly. This
distinction matters because multiple rules may target the same capability
through different actuation paths. Each device also exposes one or more
\texttt{State} nodes encoding observable properties such as \texttt{lock\_status},
\texttt{cam\_power}, \texttt{temperature\_reading}, and \texttt{motion\_detected}.
States are the primary targets of triggers and conditions, connecting the
automation layer back to the physical world.

An \texttt{EVar} (Environmental Variable) node captures room-scoped physical
quantities (temperature, humidity, illuminance, airflow, smoke concentration)
that are not directly tied to a single device but are influenced by device
actions and may trigger safety-relevant rule responses. EVars are the
mediating layer through which environment-mediated semantic conflicts propagate.

\emph{Automation layer.} A \texttt{Context} node encodes a derived home
predicate (\emph{Night}, \emph{Away}, \emph{NoMotion10m}) inferred from device
states or environmental variables. Contexts allow multiple rules to share a
high-level mode description without duplicating the underlying state logic,
and they introduce another class of indirect dependency: two rules may interact
through a shared context even if they reference entirely different physical
devices. A \texttt{Rule} node encodes a complete TCA automation, decomposed
into three structural sub-nodes: \texttt{Trigger}, \texttt{Condition}, and
\texttt{Action}. Exposing these sub-nodes explicitly---rather than embedding TCA
structure in a monolithic rule node---allows conflict-detection traversal to
reason about partial overlaps, such as two rules that share a trigger but diverge
in conditions or actions, or share a target but assign it different values.

\begin{table}[t]
  \centering
  \caption{Node types in the \textsc{SHACR} knowledge graph.}
  \label{tab:node_types}
  \scriptsize
  \setlength{\tabcolsep}{3pt}
  \renewcommand{\arraystretch}{1.1}
  \resizebox{\columnwidth}{!}{%
  \begin{tabular}{lp{6cm}}
    \toprule
    \textbf{Node Type} & \textbf{Semantic Role} \\
    \midrule
    \texttt{Building}   & Root entity containing one or more apartments. \\
    \texttt{Apartment}  & Residential unit; boundary for automation scope and deployment. \\
    \texttt{Room}       & Spatial subdivision; localizes devices, EVars, and contextual conditions. \\
    \texttt{Device}     & Physical IoT sensor or actuator installed in a room. \\
    \texttt{Capability} & Abstract operation a device can perform (e.g., power, lock, heat, record). \\
    \texttt{State}      & Observable device attribute that may change over time (e.g., lock\_status, cam\_power, motion\_detected). \\
    \texttt{EVar}       & Room-scoped ambient quantity (temperature, humidity, illuminance, airflow, smoke concentration). \\
    \texttt{Context}    & Derived predicate computed from states or EVars (e.g., \textit{Night}, \textit{Away}, \textit{EnergySavingOn}). \\
    \texttt{Rule}       & Complete TCA automation rule encoding when and how actions execute. \\
    \texttt{Trigger}    & Event or state transition that initiates rule evaluation. \\
    \texttt{Condition}  & Logical predicate that must hold for the rule to execute. \\
    \texttt{Action}     & Effect executed when the rule fires, actuating a capability or influencing a state. \\
    \bottomrule
  \end{tabular}}
\end{table}

The graph uses 12 directed, typed edges summarized in
Table~\ref{tab:Edge_Types}. The most analytically significant is
\texttt{AFFECTS}: $(\texttt{Capability}) \rightarrow (\texttt{State} \mid
\texttt{EVar})$. This edge encodes physical cause-effect relationships among
device capabilities and environmental or state variables---relationships that are
invisible to any text-based analysis. For example, the edge
$\texttt{power} \xrightarrow{\textsc{affects}} \texttt{cam\_power}$ encodes
the physical fact that switching a smart plug's power capability changes the
camera's operational state; this dependency cannot be inferred from either
rule's text. Similarly, $\texttt{run} \xrightarrow{\textsc{affects}}
\texttt{motion\_detected}$ encodes that the vacuum's movement produces sensor
activations in rooms it traverses. The \texttt{AFFECTS} edges collectively form
the physical interaction substrate over which all three conflict classes are
detected: a logical conflict requires no \texttt{AFFECTS} traversal; a semantic
conflict traverses exactly one \texttt{AFFECTS} edge to an \texttt{EVar}; a
physical conflict traverses a heterogeneous multi-hop path that may include
several \texttt{AFFECTS} edges and cross multiple node types.

\begin{table}[t]
  \centering
  \caption{Directed edge types in the \textsc{SHACR} knowledge graph.}
  \label{tab:Edge_Types}
  \scriptsize
  \renewcommand{\arraystretch}{1.1}
  \resizebox{\columnwidth}{!}{%
  \begin{tabular}{p{2.4cm} p{4.0cm} p{3.2cm}}
    \toprule
    \textbf{Edge} & \textbf{Direction} & \textbf{Interpretation} \\
    \midrule
    \texttt{HAS\_APARTMENT} & $(\texttt{Building}) \to (\texttt{Apartment})$   & Associates apartments with a building. \\
    \texttt{HAS\_ROOM}      & $(\texttt{Apartment}) \to (\texttt{Room})$        & Defines rooms within an apartment. \\
    \texttt{HAS\_DEVICE}    & $(\texttt{Room}) \to (\texttt{Device})$           & Places devices in their physical rooms. \\
    \texttt{HAS\_CAP}       & $(\texttt{Device}) \to (\texttt{Capability})$     & Specifies the operational capabilities of a device. \\
    \texttt{HAS\_STATE}     & $(\texttt{Device}) \to (\texttt{State})$          & Captures observable device attributes. \\
    \texttt{HAS\_EVAR}      & $(\texttt{Room}) \to (\texttt{EVar})$             & Associates ambient environmental variables with their rooms. \\
    \texttt{AFFECTS}        & $(\texttt{Capability}) \to (\texttt{State} \mid \texttt{EVar})$ & Encodes the physical or logical effect of actuating a capability on states or environmental variables. \\
    \texttt{HAS\_TRIGGER}   & $(\texttt{Rule}) \to (\texttt{Trigger})$          & Links a rule to its triggering event or state transition. \\
    \texttt{HAS\_CONDITION} & $(\texttt{Rule}) \to (\texttt{Condition})$        & Links a rule to its logical preconditions. \\
    \texttt{HAS\_ACTION}    & $(\texttt{Rule}) \to (\texttt{Action})$           & Links a rule to the action it executes when fired. \\
    \texttt{TARGETS}        & $(\texttt{Trigger}\mid\texttt{Cond.}\mid\texttt{Action}) \to (\texttt{State}\mid\texttt{EVar}\mid\texttt{Cap.}\mid\texttt{Context})$ & Identifies the entity observed, checked, or actuated by a TCA component. \\
    \texttt{DERIVED\_FROM}  & $(\texttt{Context}) \to (\texttt{State} \mid \texttt{EVar})$ & Records that a contextual predicate is inferred from underlying states or EVars. \\
    \bottomrule
  \end{tabular}}
  \vspace{-0.6 cm}
\end{table}

\subsection{Formal Conflict Definitions}
\label{subsec:conflict_defs}

\begin{figure}[t]
  \centering
  \includegraphics[width=0.46\textwidth]{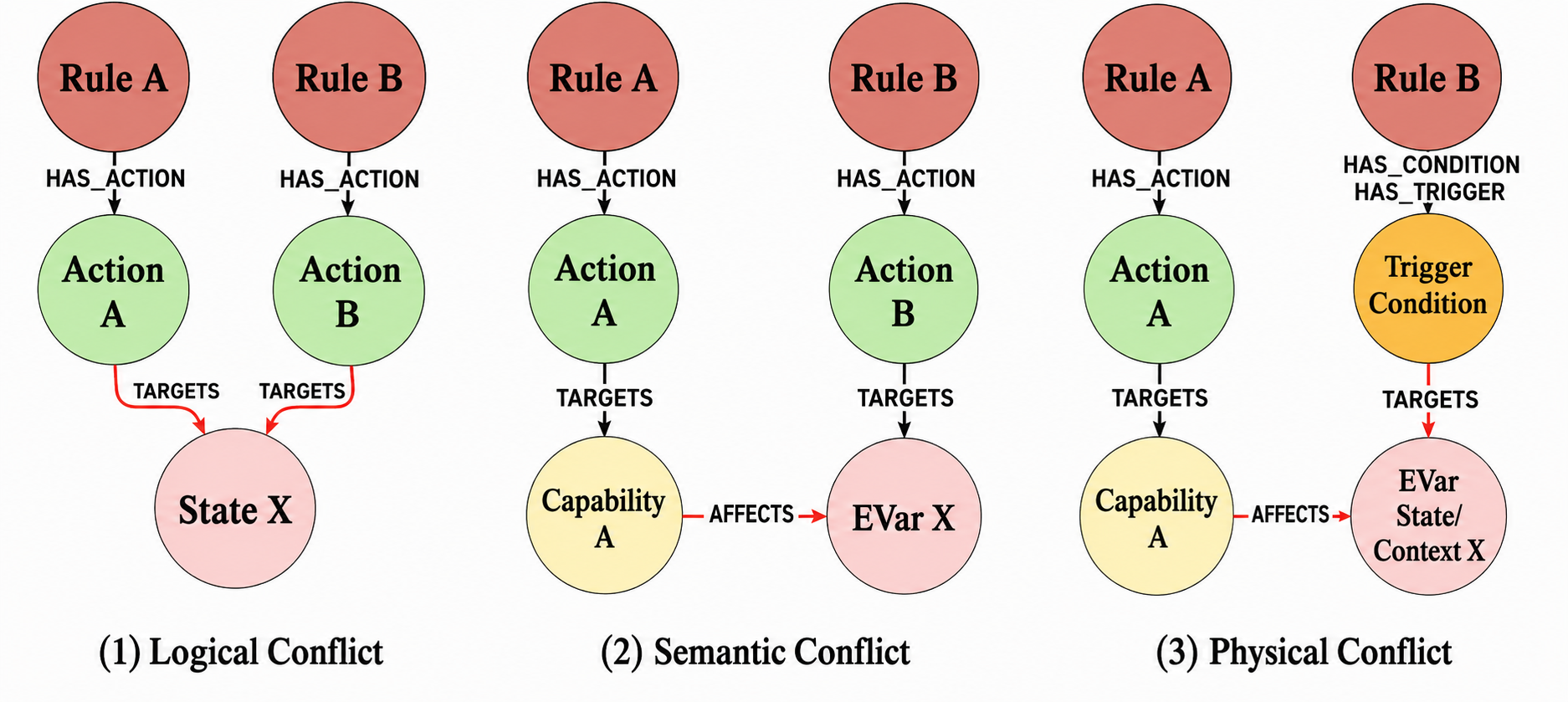}
  \caption{Structural characterization of the three conflict classes.
  (1)~\emph{Logical}: two actions target the same \texttt{State} node with
  incompatible values. (2)~\emph{Semantic}: an action propagates through a
  \texttt{Capability}$\to$\texttt{AFFECTS}$\to$\texttt{EVar} chain to activate
  a second rule's trigger or condition. (3)~\emph{Physical}: a multi-hop path
  through heterogeneous entity types links two rules without any direct
  dependency edge, composing into an unsafe global state.}
  \label{fig:C3}
\end{figure}

The following definitions formalize the three conflict classes over the directed
knowledge graph $G=(V,E,\tau_V,\tau_E,A_V,A_E)$ introduced above.
Figure~\ref{fig:C3} illustrates their structural distinction.

\newtheorem{definition}{Definition}

\begin{definition}[Logical Conflict]
A \emph{logical conflict} exists when two automation rules assign mutually
exclusive values to the same device state under overlapping execution contexts:

\begin{tcolorbox}[defbox]
\begin{equation}
\label{eq:logical}
\begin{aligned}
&\exists\, r_i, r_j, s, a_i, a_j \in V:\\
&\quad \tau_V(r_i)=\tau_V(r_j)=\texttt{Rule},\quad \tau_V(s)=\texttt{State},\\
&\quad \tau_V(a_i)=\tau_V(a_j)=\texttt{Action},\\
&\quad (r_i \xrightarrow{\texttt{HAS\_ACTION}} a_i)\wedge(r_j \xrightarrow{\texttt{HAS\_ACTION}} a_j),\\
&\quad (a_i \xrightarrow{\texttt{TARGETS}} s)\wedge(a_j \xrightarrow{\texttt{TARGETS}} s),\\
&\quad A_V(a_i).\texttt{value} \neq A_V(a_j).\texttt{value}.
\end{aligned}
\end{equation}
\end{tcolorbox}
\end{definition}

Rules $r_i$ and $r_j$ each own an \texttt{Action} sub-node, $a_i$ and $a_j$
respectively, both of which target the same \texttt{State} node $s$ via
\texttt{TARGETS}; the conflict condition is that the attribute values assigned
by $a_i$ and $a_j$ to $s$ are mutually exclusive. Logical conflicts are direct
rule-level contradictions detectable through static constraint analysis without
modeling any environmental propagation, because the shared target $s$ is
referenced explicitly in both rules' action clauses.

\begin{definition}[Semantic Conflict]
A \emph{semantic conflict} exists when the execution of one rule indirectly
activates another through shared environmental variables, forming an implicit
causal chain absent from the automation logic text:

\begin{tcolorbox}[defbox]
\begin{equation}
\label{eq:semantic}
\begin{aligned}
&\exists\, r_i, r_j, e, c, a_i, t_j \in V :\\
&\quad\tau_V(r_i)=\tau_V(r_j)=\texttt{Rule},\quad\tau_V(e)=\texttt{EVar},\\
&\quad\tau_V(c)=\texttt{Capability},\quad\tau_V(a_i)=\texttt{Action},\\
&\quad\tau_V(t_j)\in\{\texttt{Trigger},\texttt{Condition}\},\\
&\quad(r_i \xrightarrow{\texttt{HAS\_ACTION}} a_i)\wedge
     (a_i \xrightarrow{\texttt{TARGETS}} c)\wedge
     (c \xrightarrow{\texttt{AFFECTS}} e),\\
&\quad(t_j \xrightarrow{\texttt{TARGETS}} e)\wedge
     (r_j \xrightarrow{\texttt{HAS\_TRIGGER}\,\vee\,\texttt{HAS\_CONDITION}} t_j).
\end{aligned}
\end{equation}
\end{tcolorbox}
\end{definition}

where the consequent execution of $r_j$ produces an undesired or unsafe system
state. Rule $r_i$ fires an action $a_i$ that targets capability $c$; actuating
$c$ propagates through the \texttt{AFFECTS} edge to modify environmental variable
$e$; the trigger or condition sub-node $t_j$ of rule $r_j$ observes $e$ via
\texttt{TARGETS}; and $r_j$ is linked to $t_j$ through \texttt{HAS\_TRIGGER} or
\texttt{HAS\_CONDITION}. Semantic conflicts capture environment-mediated couplings
between rules that appear syntactically independent, because neither rule's text
encodes the environmental variable $e$ that mediates between them.

\begin{definition}[Physical Conflict]
A \emph{physical conflict} exists when multi-hop interactions through
heterogeneous entity types produce unsafe behavior without any direct
rule-to-rule dependency. Formally, a physical conflict exists if there is a
path $P = \langle r_i, v_1, \ldots, v_k, r_j \rangle$ in $G$ such that:

\begin{tcolorbox}[defbox]
\begin{equation}
\label{eq:physical}
\begin{aligned}
&\exists\, r_i, r_j \in V,\;
  \exists\, v_1, \ldots, v_k \in V,\;
  \exists\, P = \langle r_i, v_1, \ldots, v_k, r_j \rangle :\\
&\quad\text{(i)}\;\tau_V(r_i)=\tau_V(r_j)=\texttt{Rule};\\
&\quad\text{(ii)}\;\{v_1,\ldots,v_k\}\text{ is a heterogeneous mix of }\{\texttt{Device,}\\
&\qquad\texttt{Capability, EVar, State, Context}\};\\
&\quad\text{(iii)}\;\nexists\text{ direct dependency edge between }r_i\text{ and }r_j;\\
&\quad\text{(iv)}\;\text{the composed effects along }P\text{ violate }\Phi(G).
\end{aligned}
\end{equation}
\end{tcolorbox}
\end{definition}

Two \texttt{Rule} nodes $r_i$ and $r_j$ are connected by a path $P$ through
an intermediate sequence of nodes $v_1, \ldots, v_k$ drawn from a heterogeneous
mix of entity types; no direct dependency edge exists between $r_i$ and $r_j$
in $E$, so the interaction is entirely mediated by the intermediate path; and
the composed physical effects along $P$ violate the system safety predicate
$\Phi(G)$. Note that $c$ does not appear in this definition because physical
conflicts generalize beyond single-capability propagation to arbitrary multi-hop
chains that may traverse several \texttt{AFFECTS} edges and multiple node types.
Physical conflicts arise from latent cyber--physical interactions that are
visible only when reasoning over the combined structure of the deployment
environment, not from any individual rule's text.

\subsection{Dependability Implications of the Conflict Taxonomy}

The three conflict classes carry distinct consequences for system dependability.
Logical conflicts are primarily \emph{reliability failures}: contradictory
commands leave actuators in undefined or oscillating states, degrading system
correctness without necessarily producing immediate physical harm. Semantic
conflicts manifest as \emph{safety failures}: an implicit environmental chain
produces a hazardous system state---an overheated room, a locked exit during a
fire alarm---that the user never authorized and that cannot be diagnosed from
rule text alone. Physical conflicts are the most consequential because a single
\texttt{AFFECTS} path can simultaneously constitute a safety hazard, a security
vulnerability, and a reliability degradation through the same causal chain. The
knowledge graph substrate makes all three consequence dimensions inspectable
over the same traversal structure, which is why building that structure correctly
is the precondition for all subsequent analysis.

\section{System Design}
\label{sec:design}

\subsection{Design Rationale}
\label{subsec:rationale}

The core insight driving \textsc{SHACR}'s design is that smart home conflicts
are routinely undetected not because they are rare, but because existing
representations are insufficient to expose them. A directed knowledge graph
restores that visibility by making device relationships, environmental effects,
and rule interactions explicit, structured, and inspectable. Because dependencies
are encoded as typed graph edges, detecting conflicts becomes a matter of
traversing the structure rather than inferring it from natural language, enabling
analysis that rule-text processing cannot achieve regardless of the reasoning
method applied on top.

Within this framework, automation rules are treated not as isolated text but as
interconnected components embedded in the graph through their TCA structure.
This enables the system to reason about both direct rule interactions and
indirect multi-hop effects propagated through shared states, environmental
variables, and capability-induced changes. Building on this representation,
\textsc{SHACR} adopts a GraphRAG pipeline in which apartment-level subgraphs
are retrieved as the focused reasoning context for a given query. These subgraphs
provide a complete structural overview of the apartment's rules, devices, rooms,
and inter-entity relationships, which the LLM uses to generate precise conflict
detection, classification, and repair suggestions.

A central design decision is the choice to anchor conflict reasoning in an LLM
rather than a purely formal or rule-based verification engine. Non-LLM
approaches---SMT-based solvers, finite state machine verifiers, formal conflict
checkers---offer deterministic, auditable outputs with no inference variability.
They are computationally efficient and produce results that can be formally
proven correct within their specification. For strictly logical conflicts, where
two rules produce directly contradictory actions on a shared device state, such
approaches are well-suited. However, the conflict landscape in smart home
automation extends well beyond logical contradictions. Physical conflicts arise
from indirect environmental interactions across heterogeneous device types, and
semantic conflicts emerge from implicit user-intent mismatches that have no
formal encoding in rule syntax. These conflict types cannot be captured by
pattern matching or formal verification alone, as they require reasoning over
natural-language rule descriptions, contextual knowledge about device
capabilities, and causal chains that span multiple graph hops.

The GraphRAG-LLM design addresses this by grounding LLM inference in structured
graph relationships, reducing hallucination risk while preserving the semantic
reasoning flexibility that formal methods lack~\cite{lewis2020retrieval, Pan2025KnowledgeGraphsLLM,guu2020retrieval}. Relevant evidence in this domain is often distributed
across entities connected through semantic relations rather than surface-level
lexical similarity; graph-based retrieval enables evidence aggregation along
explicit relational paths, supporting multi-hop reasoning and improving coherence
in structured tasks~\cite{liang2025kag}. In practice, smart home knowledge is
initially distributed across heterogeneous or semi-structured sources, and rules
are authored in natural language by non-expert users without formal specification.
A graph-augmented LLM pipeline therefore facilitates the transformation of this
unstructured knowledge into structured representations that support relational
reasoning while retaining the flexibility to handle incomplete or ambiguous rule
descriptions~\cite{wang2024knowledge}. The overall design is additionally informed
by empirical findings on the challenges non-expert smart home users face in
forming accurate mental models of rule interactions and recovering from
automation failures~\cite{Huang2015SupportingMentalModel,Brackenbury2019InterpretBugsTAP, Wozniak2025ConnectingDots,Davidoff2006PrinciplesSmartHome,Caivano2019SupportingEndUsers,Koshy2021PassengerPerspectives,Zhang2022HelpingUsersDebug,Zaidi2023UserCentricConflictResolution}.

\subsection{Architecture Overview}
\label{subsec:architecture}

\begin{figure}[t]
  \centering
  \includegraphics[width=\linewidth]{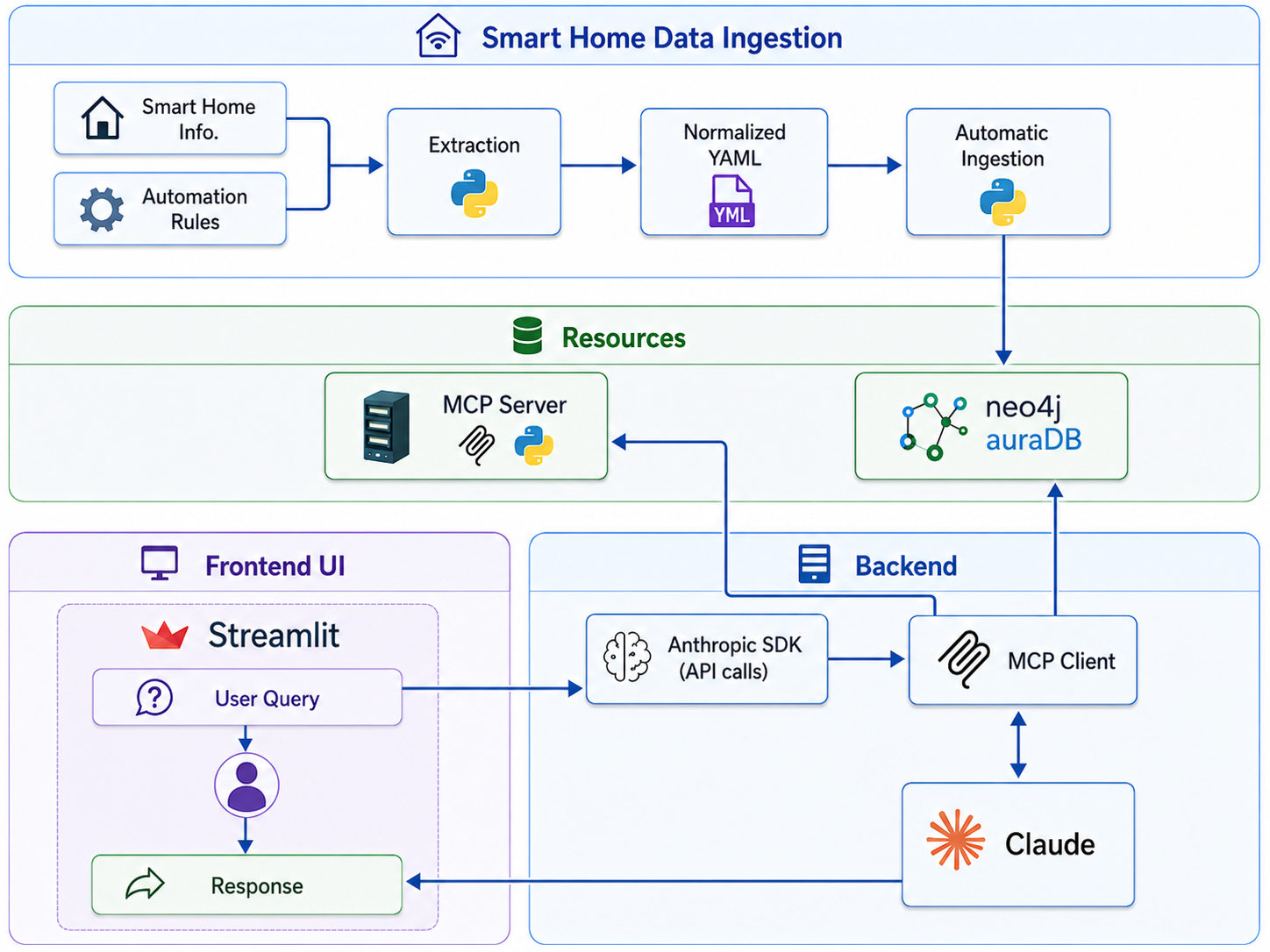}
  \caption{\textsc{SHACR} system architecture. Automation rules are ingested
  from YAML specifications and normalized into a Neo4j knowledge graph. The
  backend reasoning engine accesses the graph through two MCP server tiers. All
  knowledge graph write operations are gated behind explicit user approval in
  the Streamlit dashboard.}
  \label{fig:architecture}
  \vspace{-0.5cm}
\end{figure}

\textsc{SHACR} comprises three functional layers illustrated in Figure~\ref{fig:architecture}: (1)~a \emph{data ingestion layer} that normalizes heterogeneous rule specifications into the knowledge graph; (2)~a \emph{backend reasoning layer} comprising the Neo4j graph store, two-tier MCP server, and LLM orchestrator; and (3)~a \emph{user interface layer} that enforces human approval over all graph modifications.

\subsubsection{Data Ingestion}
\label{subsubsec:ingestion}

Smart home configurations and automation rules are provided as raw YAML
specifications authored on heterogeneous platforms such as Home Assistant and 
SmartThings. The ingestion pipeline begins with an extraction
and normalization module that parses these specifications and maps them into a
common TCA representation, resolving platform-specific syntax variations so
that the subsequent analysis is independent of the originating platform. The
normalization step also resolves device naming conventions, matches capability
descriptors to the standard node taxonomy defined in \S\ref{subsec:schema}, and
identifies implicit dependencies---such as a device's downstream power dependency
on a smart plug---that must be made explicit as \texttt{AFFECTS} edges.

Once normalized, the structured representations are ingested into the Neo4j
knowledge graph by an automated ingestion script. The script creates typed
nodes for all entities in the home (\texttt{Building}, \texttt{Apartment},
\texttt{Room}, \texttt{Device}, \texttt{Capability}, \texttt{State},
\texttt{EVar}, \texttt{Context}, \texttt{Rule}, and the TCA sub-nodes) and
instantiates the directed edge relationships defined in Table~\ref{tab:Edge_Types}.
The \texttt{AFFECTS} edges between capability nodes and their downstream state
or environmental variable nodes are populated during ingestion using a combination
of device documentation lookup and rule-text analysis, ensuring that the physical
cause-effect substrate is complete before any conflict reasoning begins. 

\subsubsection{Backend Reasoning Engine}
\label{subsubsec:backend}

The backend comprises three interacting components: the Neo4j AuraDB knowledge
graph store, two MCP server tiers, and the LLM-based reasoning module.

\textbf{Knowledge graph store.} The Neo4j AuraDB instance maintains the
structured smart home representation constructed during ingestion. It is accessed
through an encrypted connection using the \texttt{neo4j+s://} protocol, with
credentials stored as environment variables. The database serves both as the
persistent knowledge repository and as the retrieval source for the GraphRAG
pipeline. Any repair approved by the user is written back to this store through
the sole write tool in the MCP layer, ensuring that the graph reflects the
current configuration state at all times.

\textbf{MCP server layer.} The Model Context Protocol (MCP) is an open protocol
developed by Anthropic that provides a standardized interface through which the
LLM can invoke external tools and data sources at inference
time~\cite{anthropic2024mcp}. Two MCP servers operate concurrently in this
system. The first, \texttt{mcp-neo4j-cypher}, is a general-purpose open-source
server that exposes three primitive database operations (\texttt{read\_cypher},
\texttt{write\_cypher}, and \texttt{get\_schema}), launched automatically via
the \texttt{uvx} package runner using the stdio transport protocol.

\begin{figure}[t]
  \centering
  \includegraphics[width=0.8\linewidth]{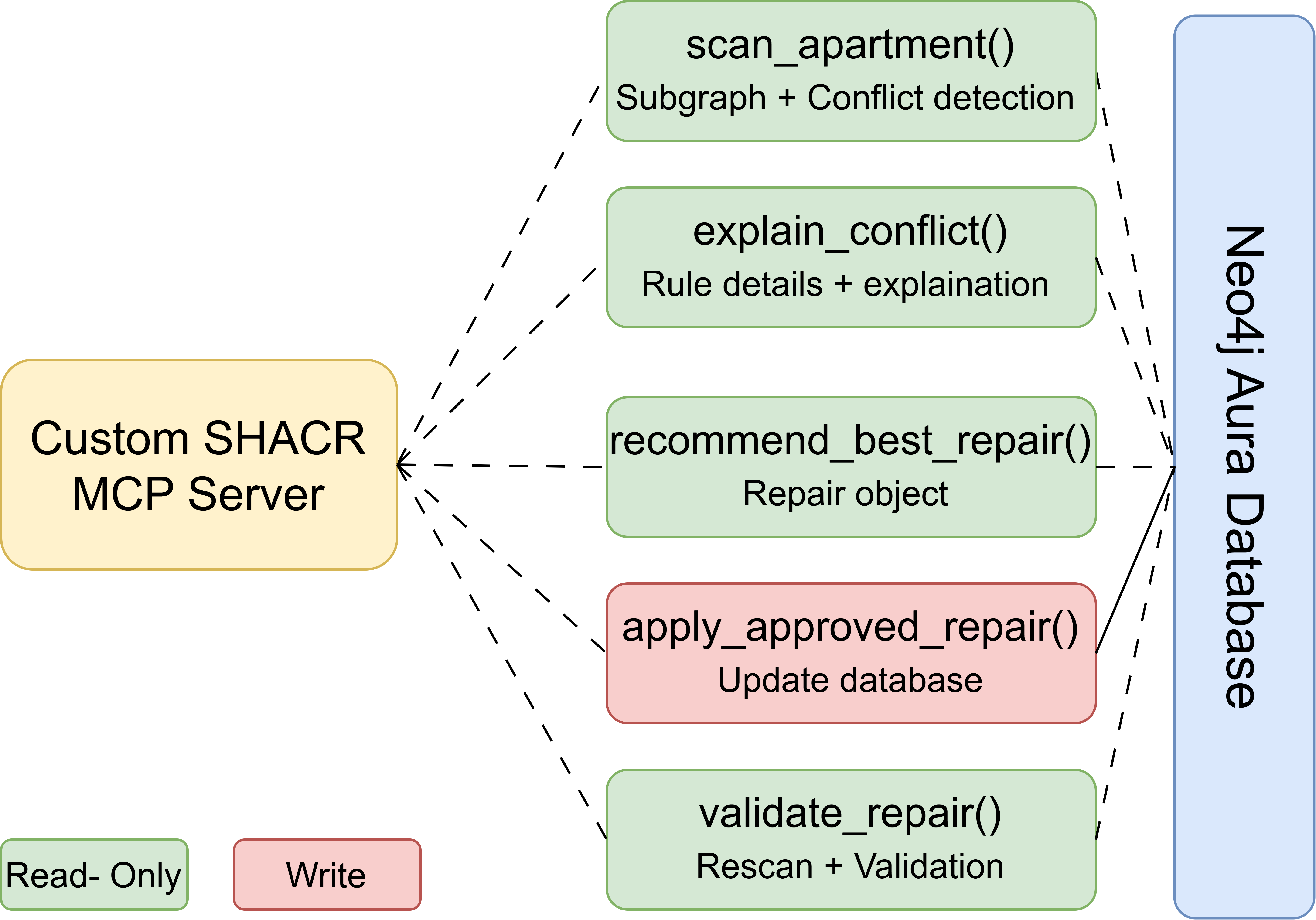}
  \caption{The five custom MCP tools forming the operational backbone of
  \textsc{SHACR}. Read-only tools are invoked autonomously by the reasoning
  engine; the single write tool requires explicit user approval before execution.}
  \label{fig:tools}
\end{figure}

The second is a custom domain-specific MCP server that exposes five
purpose-built tools (Figure~\ref{fig:tools}) to the LLM over a standardized
protocol interface. These tools collectively define the action space available
to the reasoning agent, enforce the system's safety constraints, and mediate
all interactions with the knowledge graph. Four tools are strictly read-only;
only one tool performs write operations, and it is gated behind explicit user
approval. This design prevents the LLM from autonomously modifying the knowledge
graph and ensures that all graph state changes are traceable to a specific user
decision.

\paragraph{Tool~1: \texttt{scan\_apartment}}
This tool is the entry point for every \textsc{SHACR} session. It executes two
lightweight Cypher queries against the knowledge graph and returns a merged raw
subgraph covering all rules associated with the target apartment. The first
query retrieves each rule's actions, action targets (\texttt{State},
\texttt{Capability}, or \texttt{EVar} nodes), and the devices those actions
apply to. The second query retrieves each rule's triggers, conditions, and the
full capability-to-EVar chain accessible from the rule's associated devices.
The two result sets are merged in Python by \texttt{ruleId} into a unified
\texttt{rawSubgraph} before being passed to the LLM for reasoning. The tool
also returns an \texttt{apartmentDashboard} payload containing room, device,
and rule counts for display in the Streamlit interface.

\paragraph{Tool~2: \texttt{explain\_conflict}}
This tool accepts a conflict identifier of the form \texttt{ruleAId\_ruleBId}
and constructs a structured explanation object that exposes the conflicting
trigger events, contradicting action sequences, and plain-language descriptions
of both rules. The LLM uses this object to produce a human-readable explanation
suitable for presentation to the user. This tool is particularly important for
semantic and physical conflicts, where the causal chain spans multiple graph hops
and cannot be communicated clearly without tracing the specific nodes and edges
involved. Rather than asking the user to interpret raw graph data, the system
surfaces a causally grounded narrative of why the conflict exists, directly
traceable to the home's physical configuration.

\paragraph{Tool~3: \texttt{recommend\_best\_repair}}
This is the system's planning tool. It retrieves the full TCA details of both
conflicting rules alongside the current device states of the affected devices
and proposes the best-fit repair from within a constrained set of allowed
operations. To regulate LLM behavior and minimize the risk of unsafe or
semantically inappropriate repair suggestions, the model is restricted to a
bounded action space while a corresponding set of forbidden operations is
explicitly enforced (Table~\ref{tab:llm_actions}). This ensures that all
repair proposals remain within a safe, auditable operational boundary.

\begin{table}[t]
  \centering
  \caption{Permitted and forbidden repair operations in \textsc{SHACR}.}
  \label{tab:llm_actions}
  \setlength{\tabcolsep}{4pt}
  \renewcommand{\arraystretch}{1.2}
  \footnotesize
  \begin{tabular}{@{}p{0.48\linewidth} p{0.48\linewidth}@{}}
    \toprule
    \textbf{Allowed Operations} & \textbf{Forbidden Operations} \\
    \midrule
    Modify trigger conditions (\texttt{modify\_trigger}) &
    Delete rules, devices, rooms, or apartments \\
    Add mutual-exclusion conditions (\texttt{add\_condition}) &
    Invent new graph entities \\
    Remove redundant or conflicting conditions (\texttt{remove\_condition}) &
    Remove physical topology edges (\texttt{HAS\_DEVICE}, \texttt{HAS\_ROOM}, \texttt{HAS\_STATE}) \\
    Refine existing conditions for specificity (\texttt{refine\_condition}) &
    Modify graph structure beyond rule-level constraints \\
    Modify conflicting actions to resolve direct contradictions (\texttt{modify\_action}) &
    Introduce semantically arbitrary repairs \\
    Adjust rule priority to control execution order (\texttt{add\_priority}) &
    Execute write operations without user approval \\
    \bottomrule
  \end{tabular}
\end{table}

\paragraph{Tool~4: \texttt{apply\_approved\_repair}}
This is the sole write tool in the system and the only point at which the
knowledge graph is modified. It accepts the structured repair object produced
by \texttt{recommend\_best\_repair} and executes the corresponding Cypher write
against Neo4j. The tool validates that the repair object contains both a target
rule and a specified operation before executing, and returns a structured success
or failure response. Currently implemented operations include
\texttt{add\_condition}, which creates a new \texttt{Condition} node with a
mutual-exclusion constraint. This tool cannot be invoked autonomously by the
LLM: explicit user approval is a hard architectural requirement, not a soft
preference, ensuring that all knowledge graph modifications are supervised.

\paragraph{Tool~5: \texttt{validate\_repair}}
This tool re-executes a targeted conflict-detection query scoped to the two
rules involved in the original conflict, determines whether the logical
contradiction between their actions on the shared device still exists, and
returns a boolean \texttt{isResolved} flag alongside the current condition
state of both rules. If the conflict persists, the tool reports the remaining
conflict count so that the reasoning engine can generate an alternative repair.
Together with \texttt{apply\_approved\_repair}, this tool implements the
Scan--Approve--Execute--Verify cycle that forms the operational core of
\textsc{SHACR}'s conflict resolution pipeline.

\textbf{Reasoning module.} The LLM operates in a semi-autonomous capacity as
the system's reasoning orchestrator. It is connected to both MCP servers through
the Anthropic SDK and discovers all available tools at inference time. Upon
receiving a user query, it selects and invokes the appropriate MCP tool,
processes the returned structured subgraph data, performs conflict analysis, and
generates a structured response comprising detected conflicts and repair
recommendations. This positions the LLM as an active orchestrator rather than a
passive text generator, enabling multi-step reasoning pipelines that interleave
graph retrieval, conflict analysis, and repair generation within a single
inference cycle. To ensure bounded and auditable behavior, the LLM's action
space is constrained by the MCP tool definitions and the system prompt, and no
write operations may be executed without explicit user approval.

\subsubsection{User Interface}
\label{subsubsec:ui}

The Streamlit frontend provides an interactive dashboard through which users
initiate and monitor the conflict resolution workflow. A user submits an
apartment identifier to trigger a full apartment scan, which invokes the
reasoning pipeline with the active MCP configuration. The returned response is
rendered as structured summary metrics, a list of detected conflicts with
graph-grounded causal explanations, and corresponding repair recommendations. A
mandatory approval step is enforced before any repair action is applied to the
knowledge graph: the system surfaces the specific repair proposal (what change
will be made, which rules are affected, and what the expected outcome is) and
requires an explicit confirmation before \texttt{apply\_approved\_repair} is
invoked. This human-in-the-loop mechanism ensures that the system operates
as a decision-support tool, with final authority over the home configuration
remaining with the user.

\subsection{Failure Conditions and Scope}
\label{subsec:failure_conditions}

\textsc{SHACR} targets pre-deployment conflict detection in a multi-platform
smart building environment and does not assume adversarial occupants. It is,
however, designed to be robust to two categories of imperfect input that arise
naturally in real deployments.

\textbf{Imperfect rule descriptions.} Automation rules may contain misleading,
incomplete, ambiguous, or semantically inconsistent descriptions. These
imperfections may arise from user misconfiguration, cross-platform semantic
mismatches, automatically generated rule templates, or inadequately described
device capabilities. When rule descriptions do not accurately reflect the
intended semantics of the deployment, the constructed knowledge graph may
become inconsistent or incomplete. This can result in incorrect reasoning about
device interactions, failure to detect unsafe physical conflicts, and unsafe
repair suggestions. The framework addresses this risk through the formal graph
schema: the normalization step during ingestion maps rule descriptions against
the typed node and edge taxonomy, and mismatches are surfaced as graph
construction anomalies rather than silently propagated into the reasoning
pipeline.

\textbf{LLM-specific reasoning limitations.} Because the framework uses an LLM
for rule interpretation, conflict explanation, and repair suggestion, it inherits
the inference variability of generative models. The LLM may produce outputs that
appear plausible yet contradict the underlying knowledge graph, or exhibit
overconfidence on borderline classification cases. The framework mitigates these
risks through four architectural measures: (1)~all LLM reasoning is grounded in
Cypher-retrieved subgraphs rather than free-form text; (2)~deterministic graph
traversal provides an independent post-classification verification step;
(3)~repairs are restricted to a bounded, semantically validated operation set;
and (4)~all write operations require explicit user approval before execution.

\section{Prototype Implementation}
\label{sec:impl}

\subsection{System Configuration}
\label{subsec:system_config}

The \textsc{SHACR} stack integrates four principal components: a Neo4j AuraDB
graph database, an MCP server layer, an LLM reasoning orchestrator, and a
Streamlit web interface. Table~\ref{tab:components} lists the specific
technologies, versions, and roles of each component.

\begin{table}[t]
  \centering
  \caption{System components, versions, and roles.}
  \label{tab:components}
  \setlength{\tabcolsep}{3pt}
  \renewcommand{\arraystretch}{1.05}
  \footnotesize
  \begin{tabular}{@{}l l p{2.9cm}@{}}
    \toprule
    \textbf{Component} & \textbf{Version} & \textbf{Role} \\
    \midrule
    Neo4j AuraDB      & 5.26 LTS (cloud) & Knowledge graph store \\
    mcp-neo4j-cypher  & 0.6.0            & General-purpose MCP bridge to Neo4j \\
    Claude Sonnet~4   & claude-sonnet-4-20250514 & Reasoning orchestrator \\
    Custom MCP Server & Python 3.10+     & Domain-specific tool layer \\
    Streamlit         & 1.45.1           & User dashboard interface \\
    Anthropic SDK     & 0.107.1          & Claude client; drives MCP server \\
    \bottomrule
  \end{tabular}
\end{table}

Both MCP servers are registered in the host configuration file and instantiated as subprocesses at application startup. The reasoning module discovers all available tools at inference time and selects the appropriate tool based on the active user query and system prompt instructions. Connection credentials for Neo4j are stored as environment variables and are never embedded in source code, in accordance with standard security practices.

\subsection{Runtime Workflow}
\label{subsec:runtime}

\begin{figure}[t]
  \centering
  \includegraphics[width=0.75\linewidth]{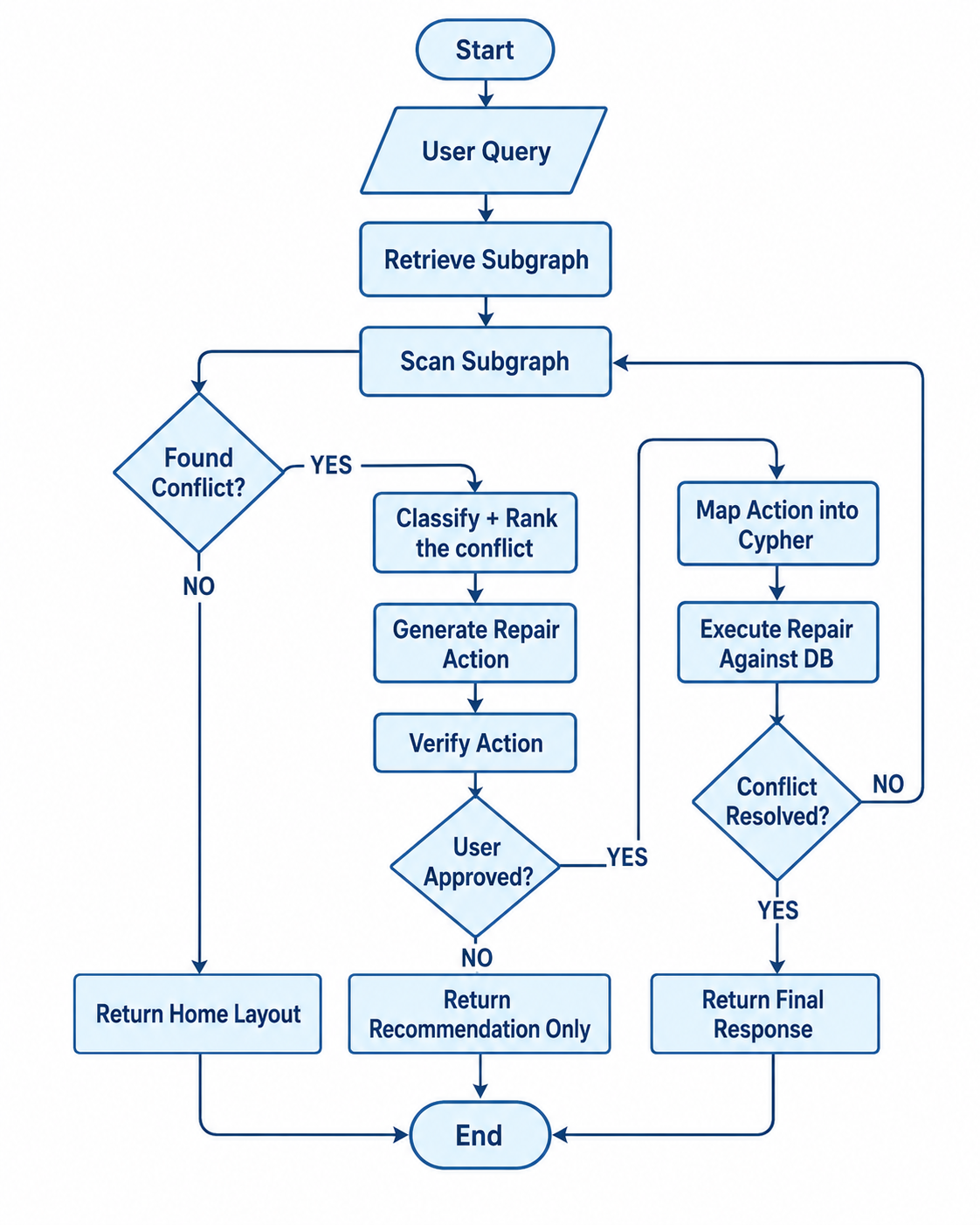}
  \caption{\textsc{SHACR} runtime workflow. Subgraph retrieval feeds conflict
  detection and classification. Each detected conflict yields a causal explanation
  and a constrained repair recommendation presented to the user. If approved,
  the repair is applied and re-verified; if the conflict persists, an alternative
  repair is generated. The workflow terminates when all conflicts are resolved or
  no valid repair can be produced.}
  \label{fig:flowchart}
\end{figure}

Figure~\ref{fig:flowchart} illustrates the end-to-end operational workflow.
The reasoning engine interprets the user's apartment-scan query, invokes
\texttt{scan\_apartment} to retrieve the relevant subgraph, and analyzes all
rule pairs within the subgraph for conflicts across the three defined classes.
For each detected conflict, the engine invokes \texttt{explain\_conflict} to
construct a causally grounded explanation, then invokes
\texttt{recommend\_best\_repair} to propose a repair from within the constrained
action space. The explanation and repair recommendation are rendered in the
Streamlit dashboard, where the user reviews the proposed change and decides
whether to approve it.

If approved, the repair is applied through \texttt{apply\_approved\_repair} and
immediately re-verified through \texttt{validate\_repair}. If the validation
confirms resolution, the conflict is marked as resolved and the workflow proceeds
to the next detected conflict. If the validation finds that the conflict persists
despite the applied repair, the system reports the unresolved state and generates
an alternative repair recommendation. The workflow terminates when either all
detected conflicts have been resolved or no valid repair can be generated within
the allowed operation set. This bounded termination condition prevents the
system from entering unbounded repair loops and ensures that the user is
informed of any conflicts that require manual intervention.

Algorithm~\ref{alg:shrag_workflow} formalizes the internal logic of the
detection and repair pipeline.

\begin{algorithm}[t]
  \caption{\textsc{SHACR} Conflict Detection and Safe Repair}
  \label{alg:shrag_workflow}
  \begin{algorithmic}[1]
    \Require Apartment identifier $\mathit{apt\_id}$
    \Ensure Classified conflict reports $C$, causal explanations,
            approved repairs, and validation outcomes
    \State Connect to Neo4j knowledge graph via two-tier MCP server
    \State $\mathit{subgraph} \gets$
           \Call{ScanApartment}{$\mathit{apt\_id}$}
           \Comment{Retrieve merged rules, devices, states, EVars, capabilities}
    \State $C \gets$ \Call{LLMDetect}{$\mathit{subgraph}$}
           \Comment{LLM identifies all conflicting rule pairs}
    \For{each $\mathit{cf} \in C$}
      \State $\mathit{cf}.\mathit{type} \gets$
             \Call{LLMClassify}{$\mathit{cf},\, \mathit{subgraph}$}
             \Comment{Classify as \textsc{Logical}, \textsc{Semantic},
                      or \textsc{Physical}}
    \EndFor
    \For{each $\mathit{cf} \in C$}
      \State $\mathit{expl} \gets$
             \Call{ExplainConflict}{$\mathit{cf}.\mathit{id}$}
             \Comment{Trace causal path through graph entities}
      \State $\mathit{repair} \gets$
             \Call{RecommendBestRepair}{$\mathit{cf}.\mathit{id}$}
             \Comment{Select from constrained action space}
      \State Present $\mathit{expl}$ and $\mathit{repair}$ to user
      \If{user approves $\mathit{repair}$}
        \State \Call{ApplyApprovedRepair}{$\mathit{repair}$}
               \Comment{Updates live knowledge graph}
        \State $\mathit{resolved} \gets$
               \Call{ValidateRepair}{$\mathit{cf}.\mathit{id}$}
        \If{$\mathit{resolved}$}
          \State Mark $\mathit{cf}$ as resolved in $C$
        \Else
          \State $\mathit{repair} \gets$
                 \Call{RecommendBestRepair}{$\mathit{cf}.\mathit{id}$}
                 \Comment{Generate alternative repair}
          \State Present alternative to user; repeat approval step
        \EndIf
      \EndIf
    \EndFor
    \State \Return $C$, explanations, repair outcomes,
           validation results
  \end{algorithmic}
\end{algorithm}

Arriving at the final retrieval design required three successive engineering
iterations. Each phase exposed a fundamental limitation that motivated the next,
and the progression from a single-query design to a split retrieval architecture
reflects the practical constraints of operating on a cloud-hosted graph database
at non-trivial scale.

\textbf{Phase~1: Single combined retrieval query.} The initial design attempted
to capture the complete apartment subgraph using a single Cypher query that
traversed rules, actions, triggers, conditions, devices, capabilities, and
environmental variables in a single execution pipeline. This design failed in
practice. The large number of chained traversal operations caused combinatorial
path expansion and consistently triggered query timeouts on Neo4j AuraDB,
particularly for apartments with larger rule sets and richer device relationship
graphs. As a result, this approach was unable to return reliable results and
was abandoned.

\textbf{Phase~2: Hardcoded conflict detection in Cypher (V1).} The second
iteration shifted strategy from raw subgraph retrieval toward embedding conflict
detection logic directly in Cypher. Rather than retrieving the full graph context
for the LLM to reason over, this approach compared rule actions pairwise on shared
devices within the query and returned pre-identified conflicting rule pairs.
This phase resolved the timeout issue and produced stable execution performance.
However, it introduced a more fundamental architectural limitation: conflict
detection was now performed by the Cypher query layer rather than by the LLM.
The detection logic was confined to direct action contradictions on the same
device state, which corresponds only to logical conflicts. Semantic conflicts
requiring causal reasoning through
\texttt{Capability}$\to$\texttt{AFFECTS}$\to$\texttt{EVar} chains and physical
conflicts involving indirect multi-hop interactions across heterogeneous entity
types were structurally undetectable under this design.

\textbf{Phase~3: Split two-query retrieval (V2, final).} The final design
fully decouples retrieval from conflict detection. Rather than embedding
detection logic into Cypher, the system retrieves the raw graph structure and
delegates all conflict reasoning to the LLM. This decoupling is achieved by
splitting the single combined query into two lightweight parallel queries.
The first query focuses on rule actions, action targets, and associated devices.
The second retrieves triggers, conditions, and the capability-to-EVar
relationships reachable through the devices involved in each rule. The two
result sets are merged in Python using \texttt{ruleId} as the join key,
producing a unified \texttt{rawSubgraph} structure that is passed to the LLM
for reasoning. Each individual query is sufficiently narrow to avoid AuraDB
timeout thresholds; together, they expose the complete contextual neighborhood
required for reasoning over all three conflict classes.

Table~\ref{tab:retrieval_trials} summarizes the progression across the three
engineering phases.

\begin{table}[t]
  \centering
  \caption{Retrieval design evolution across engineering phases.}
  \label{tab:retrieval_trials}
  \renewcommand{\arraystretch}{1.15}
  \scriptsize
  \begin{tabular}{p{0.7cm} p{2.4cm} p{3.6cm}}
    \toprule
    \textbf{Phase} & \textbf{Approach} & \textbf{Outcome} \\
    \midrule
    P1 & Single combined subgraph query &
    Repeated AuraDB timeouts due to combinatorial path expansion; unreliable execution. \\
    P2 (V1) & Hardcoded pairwise conflict detection in Cypher &
    Execution stable; detection limited to logical conflicts; semantic and physical classes structurally undetectable. \\
    P3 (V2) & Split two-query retrieval with Python merge; full LLM reasoning &
    Reliable within AuraDB timeout thresholds; complete graph context exposed for all three conflict classes. \\
    \bottomrule
  \end{tabular}
\end{table}

\section{Evaluation}
\label{sec:eval}

We evaluate \textsc{SHACR} to answer four questions: (i)~how accurately it
detects and classifies the three conflict classes; (ii)~whether lightweight
few-shot calibration improves detection, and whether that improvement is
specific to the graph-grounded system; (iii)~how much of the performance is
attributable to the knowledge graph rather than to the underlying LLM; and
(iv)~how \textsc{SHACR} compares against established prior-work detectors.

\subsection{Testbed and Setup}
\label{subsec:testbed}

To our knowledge, no publicly available benchmark exists for large-scale smart
building automation conflict detection. We therefore constructed a controlled
testbed of 70 apartments in a simulated smart building that spanned family,
bachelor, and elderly household types, capturing diverse device inventories,
lifestyle patterns, and automation complexity. Automation rules were drawn from
publicly available repositories and official platform documentation for
SmartThings and Home Assistant, providing realistic user-authored patterns
across heterogeneous IoT ecosystems \cite{tomwaldnz_ha_examples, ha_yaml_docs, smartthings_rules_api}. The rules collected were normalized into a
unified YAML schema, mapped to the building layout, and ingested following the
schema in \S\ref{subsec:schema}; where necessary, the rules were adapted to ensure structural consistency with the knowledge graph model while preserving their
original semantics.\footnote{The complete testbed---the 70 apartment YAML
specifications, the knowledge-graph ingestion scripts, and the ground-truth
conflict labels---is publicly available at \url{https://github.com/Aljoby/SHACR}.} 

Table~\ref{tab:combined_stats_eval} summarizes the testbed statistics and the
distribution of ground-truth conflict, and Table~\ref{tab:device_inventory} lists
the device inventory. The distribution is deliberately sparse and
representative of real deployments: 39 of 70 apartments contain at least one
conflicted rule and 31 are conflict-free, and among the conflicted apartments 27
contain only one or two conflicts, while 12 contain three or more, reflecting the
predominance of isolated rather than systemic conflicts at household scale. At
the rule level, 88 of 203 rules (43\%) participate in a conflict, partitioned
into 38 logical, 31 semantic, and 19 physical conflicts. Logical conflicts are
the most common and physical conflicts---those mediated by multi-hop device
topology---the rarest, consistent with deep dependency chains being less frequent
but more consequential. This sparsity makes precision as important as recall because over-flagging imposes a real burden on users within a predominantly clean rule set.

\begin{table}[t]
  \centering
  \caption{Smart building testbed statistics and ground-truth conflict
  distribution ($n=203$ rules).}
  \label{tab:combined_stats_eval}
  \setlength{\tabcolsep}{6pt}
  \renewcommand{\arraystretch}{1.2}
  \begin{tabular}{@{}lc@{}}
    \toprule
    \textbf{Metric} & \textbf{Value} \\
    \midrule
    Total apartments                   & 70 \\
    Conflicted apartments              & 39 \\
    Clean apartments                   & 31 \\
    Apartments with 1--2 conflicts     & 27 \\
    Apartments with $\geq$3 conflicts  & 12 \\
    \midrule
    Total rules                        & 203 \\
    Conflicted rules                   & 88 \\
    Clean rules                        & 115 \\
    Conflict scenarios                 & 63 \\
    \midrule
    \multicolumn{2}{@{}l}{\textit{Conflict distribution by type}} \\
    \rowcolor{shragrow}
    \quad Logical conflicts            & 38 \\
    \rowcolor{shragrow}
    \quad Semantic conflicts           & 31 \\
    \rowcolor{shragrow}
    \quad Physical conflicts           & 19 \\
    \bottomrule
  \end{tabular}
\end{table}

\begin{table}[t]
  \centering
  \caption{IoT device inventory across the testbed.}
  \label{tab:device_inventory}
  \setlength{\tabcolsep}{6pt}
  \renewcommand{\arraystretch}{1.15}
  \footnotesize
  \begin{tabular}{@{}lc@{}}
    \toprule
    \textbf{Device Category} & \textbf{Count} \\
    \midrule
    Motion sensors                        & 28 \\
    Smart lighting                        & 24 \\
    Smart cameras (indoor/entry/security) & 14 \\
    Robot vacuum cleaners                 & 12 \\
    Door/entry sensors                    & 12 \\
    Heaters                               & 10 \\
    Smart door locks                      & 10 \\
    Temperature sensors                   &  9 \\
    Safety and alarm systems              &  9 \\
    Nursery and elderly care devices      &  8 \\
    HVAC and air conditioning units       &  8 \\
    Smart windows and actuators           &  7 \\
    Smart lamps                           &  7 \\
    Smoke sensors                         &  7 \\
    Air quality sensors                   &  6 \\
    Exhaust and ventilation fans          &  6 \\
    Leak detection sensors                &  5 \\
    Water control valves                  &  5 \\
    Humidity sensors                      &  3 \\
    \bottomrule
  \end{tabular}
\end{table}

We compare \textbf{full \textsc{SHACR}} (the complete framework comprising the
knowledge graph, GraphRAG retrieval, LLM-based reasoning, and executable repair)
against two complementary families of baselines. The first is a graph-free
\emph{ablation} we call the \textbf{YAML-only baseline}, in which the identical
LLM reasoning step runs on raw YAML rule text \emph{without} the knowledge graph, evaluated with both Claude Sonnet~4 (\texttt{claude-sonnet-4-20250514}) and Google Gemini (\texttt{gemini-2.5-flash}) so that the contribution of graph grounding is separated from raw model capability. Full decoding parameters appear in the
supplementary material. Because full
\textsc{SHACR} and the YAML-only Claude baseline use the identical underlying
model, their difference isolates the effect of the knowledge graph. The second
family comprises two established rule-centric detectors from the prior
literature, \textbf{IoTC$^2$}~\cite{al2019iotc} and
\textbf{VISCR}~\cite{nagendra2019viscr}, which bound what is achievable from rule
text alone without a structured physical model (\S\ref{subsec:baseline_comparison}).
Detection is scored at the rule level using precision, recall, F1, and accuracy,
following prior IoT conflict-detection
work~\cite{kuang2025safe,chen2024tapchecker,li2024seiot,ding2018safety}.
\begin{table}[t]
  \centering
  \caption{Rule-level confusion matrix, full \textsc{SHACR} ($n=203$).}
  \label{tab:confusion_matrix}
  \setlength{\tabcolsep}{8pt}
  \renewcommand{\arraystretch}{1.3}
  \begin{tabular}{@{}lcc@{}}
    \toprule
    & \textbf{Pred.\ Conflict} & \textbf{Pred.\ Clean} \\
    \midrule
    \textbf{Actual Conflict} & 70 \ (TP) & 18 \ (FN) \\
    \textbf{Actual Clean}    & 20 \ (FP) & 95 \ (TN) \\
    \bottomrule
  \end{tabular}
\end{table}

\begin{table}[t]
  \centering
  \caption{Overall conflict detection metrics, full \textsc{SHACR} (zero-shot).}
  \label{tab:metrics}
  \setlength{\tabcolsep}{6pt}
  \renewcommand{\arraystretch}{1.2}
  \begin{tabular}{@{}lc@{}}
    \toprule
    \textbf{Metric} & \textbf{Value} \\
    \midrule
    Precision              & 0.78 \\
    Recall                 & 0.80 \\
    F1-score               & 0.79 \\
    Accuracy               & 0.81 \\
    Micro-F1 (multi-class) & 0.78 \\
    Specificity            & 0.83 \\
    \bottomrule
  \end{tabular}
\end{table}

\begin{table}[H]
  \centering
  \caption{Per-conflict-type detection breakdown, full \textsc{SHACR}
  (conflict-instance level).}
  \label{tab:per_type_performance}
  \setlength{\tabcolsep}{5pt}
  \renewcommand{\arraystretch}{1.25}
  \begin{tabular}{@{}lcccccc@{}}
    \toprule
    \textbf{Type} & \textbf{TP} & \textbf{FP} & \textbf{FN} &
    \textbf{Prec.} & \textbf{Rec.} & \textbf{F1} \\
    \midrule
    Logical             & 34 & 18 & 4  & 0.65 & 0.89 & 0.75 \\
    Semantic            & 16 & 8  & 15 & 0.67 & 0.52 & 0.58 \\
    Physical            & 20 & 6  & 6  & 0.77 & 0.77 & 0.77 \\
    \midrule
    \textbf{Micro Avg.} & 70 & 32 & 25 & 0.69 & 0.74 & 0.71 \\
    \bottomrule
  \end{tabular}
\end{table}
\subsection{Conflict Detection Across All Three Classes}
\label{subsec:detection}

We first characterize full \textsc{SHACR} in its zero-shot configuration.
Tables~\ref{tab:confusion_matrix} and~\ref{tab:metrics} report overall rule-level
performance: precision 0.78, recall 0.80, F1 0.79, and accuracy 0.81, with 70
true positives, 18 false negatives, 20 false positives, and 95 true negatives.
The low false-negative count matters in safety-critical settings, where a missed
conflict is far costlier than a conservative over-flag, and the specificity of
0.83 confirms that the system separates conflicted from clean rules well despite
the sparse class balance. The micro-averaged F1 of 0.78 across the three classes
indicates that aggregate per-class performance is balanced, even though the
classes differ markedly in difficulty.

Table~\ref{tab:per_type_performance} breaks down detection performance by conflict class. Unlike the rule-level metrics in Table~\ref{tab:confusion_matrix}, these counts are aggregated at the conflict-instance level. Because a single rule can participate in multiple distinct conflict pairs, it is counted once in the rule-level matrix but enumerated per pair here; consequently, the total false positive (FP) and false negative (FN) counts differ between the two tables. The asymmetry across classes is
structurally meaningful. Logical conflicts are detected most reliably (recall
0.89) because a direct contradiction between two actions on a shared state node is
surfaced in a single graph traversal and needs no multi-hop reasoning. Physical
conflicts---entirely invisible to text-based methods---are detected with balanced
precision and recall (both 0.77), confirming that the explicit multi-hop
\texttt{AFFECTS} paths encoded in the graph make topology-mediated interactions
tractable. Semantic conflicts remain the hardest (recall 0.52, F1 0.58):
detecting them requires the model to follow chained \texttt{AFFECTS} edges through
\texttt{EVar} nodes and judge that the resulting environmental change constitutes
an undesired interaction. The semantic false negatives, which account for the
majority of all missed detections and concentrate in apartments with longer
environmental dependency chains, point to \texttt{EVar} modeling completeness as
the main lever for closing this gap. Crucially, even zero-shot \textsc{SHACR}
detects all three classes---including the physical class that rule-centric and
text-only methods cannot reach at all (\S\ref{subsec:baseline_comparison}).

\subsection{Effect of Few-Shot Prompt Calibration}
\label{subsec:fewshot}

The zero-shot configuration exhibits two systematic error modes: over-detection
of structurally similar but benign rule pairs (false positives) and missed
multi-hop semantic chains (false negatives). We ask whether lightweight few-shot
calibration---annotated examples injected into the system prompt, with no change
to the graph or retrieval pipeline---can address both.
Table~\ref{tab:fewshot_comparison} reports three configurations for full
\textsc{SHACR} alongside the two YAML-only (graph-free) LLM models.

\begin{table*}[t]
  \centering
  \caption{Few-shot prompt calibration across three configurations, comparing
  full \textsc{SHACR} (Claude Sonnet~4 + knowledge graph) against the YAML-only
  Claude Sonnet~4 and Google Gemini baselines ($n=203$, TP+FN$=88$, TN+FP$=115$).
  \textsc{SHACR} (highlighted) leads on F1 in every configuration and reaches
  0.95 with balanced examples, while few-shot calibration barely moves the
  graph-free baselines.}
  \label{tab:fewshot_comparison}
  \footnotesize
  \renewcommand{\arraystretch}{1.3}
  \begin{tabular*}{\textwidth}{@{\extracolsep{\fill}}lcccccccc@{}}
    \toprule
    \textbf{Method} & \textbf{Acc.} & \textbf{Prec.} & \textbf{Rec.} &
    \textbf{F1} & \textbf{TP} & \textbf{TN} & \textbf{FP} & \textbf{FN} \\
    \midrule
    \multicolumn{9}{@{}l}{\textit{Exp.~1: Zero-shot}} \\
    \rowcolor{shragrow}
    \textbf{SHACR (Claude + KG)}  & \textbf{0.81} & \textbf{0.78} & \textbf{0.80} & \textbf{0.79} & 70 & 95  & 20 & 18 \\
    Claude Sonnet~4 (YAML-only)   & 0.70 & 0.75 & 0.48 & 0.59 & 42 & 101 & 14 & 46 \\
    Google Gemini (YAML-only)     & 0.72 & 0.70 & 0.64 & 0.67 & 56 & 91  & 24 & 32 \\
    \midrule
    \multicolumn{9}{@{}l}{\textit{Exp.~2: Few-shot (3 conflict examples)}} \\
    \rowcolor{shragrow}
    \textbf{SHACR (Claude + KG)}  & \textbf{0.86} & \textbf{0.80} & \textbf{0.90} & \textbf{0.84} & 79 & 95  & 20 &  9 \\
    Claude Sonnet~4 (YAML-only)   & 0.76 & 0.74 & 0.70 & 0.72 & 62 & 93  & 22 & 26 \\
    Google Gemini (YAML-only)     & 0.72 & 0.70 & 0.64 & 0.67 & 56 & 91  & 24 & 32 \\
    \midrule
    \multicolumn{9}{@{}l}{\textit{Exp.~3: Few-shot (7 balanced examples)}} \\
    \rowcolor{shragrow}
    \textbf{SHACR (Claude + KG)}  & \textbf{0.96} & \textbf{1.00} & \textbf{0.90} & \textbf{0.95} & 79 & 115 &  0 &  9 \\
    Claude Sonnet~4 (YAML-only)   & 0.73 & 0.74 & 0.60 & 0.66 & 53 & 96  & 19 & 35 \\
    Google Gemini (YAML-only)     & 0.75 & 0.74 & 0.66 & 0.70 & 58 & 95  & 20 & 30 \\
    \bottomrule
  \end{tabular*}
\end{table*}

For full \textsc{SHACR}, calibration helps substantially and monotonically.
Adding three conflict examples---one per class (Exp.~2)---raises recall from 0.80
to 0.90, dropping false negatives from 18 to 9, by anchoring the model to the
\texttt{AFFECTS}-chain patterns that signal genuine multi-hop conflicts; but
supplying only positive examples leaves false positives at 20, the expected
label-distribution bias of one-sided
prompting~\cite{beltagy2022fewshot,brown2020language}. Adding four clean examples
that target the specific false-positive patterns (Exp.~3) corrects this directly:
precision rises to 1.00 as false positives fall to 0 while recall holds at 0.90,
lifting F1 from 0.79 to \textbf{0.95} and accuracy to 0.96. Graph retrieval
supplies the structured evidence; the few-shot examples calibrate how the model
reasons over it---complementary mechanisms addressing distinct failure modes. The
perfect precision in Exp.~3 reflects the tight alignment between the examples and
the controlled testbed's conflict patterns and is likely optimistic in the wild;
the underlying mechanism, however, is general.

The same calibration tells a sharply different story for the graph-free models,
and that contrast is itself a result. Few-shot prompting barely moves them:
YAML-only Claude improves only from F1 0.59 to a peak of 0.72 and then
\emph{regresses} to 0.66 under the balanced examples, while Gemini stays
essentially flat (0.67 to 0.70). Notably, the balanced examples that drive
\textsc{SHACR} to perfect precision actively \emph{reduce} YAML-only Claude's
recall, from 0.70 to 0.60. The reason is structural. Few-shot examples teach a
decision rule, but a rule can be applied only if the features it references are
present. Distinguishing a genuine conflict from a benign look-alike requires
\texttt{EVar} ranges, \texttt{AFFECTS} semantics, and context-mode
exclusivity---information that lives in the graph, not in rule text. \textsc{SHACR} has those features, so a decision criterion such as ``a structurally similar pair is clean when its \texttt{AFFECTS} paths do not compose into a hazard'' becomes a rule the model can verify. Before flagging a pair, it queries the graph to confirm whether the shared \texttt{EVar} or composing \texttt{AFFECTS} path exists, turning each example into a checkable decision aligned with ground truth, and precision climbs to 1.00. The YAML-only models receive the same decision criterion but cannot verify it; they approximate ``shared \texttt{EVar}'' from surface similarity in device names and rule text, an unreliable proxy, so the same examples inject noise and the models grow uniformly more cautious, shedding true positives with the false ones.
Calibration thus pays off only in the presence of the graph, demonstrating that prompt engineering is not a substitute for the structured substrate but rather a complement to it.

\subsection{Contribution of the Knowledge Graph}
\label{subsec:ablation}

The configurations in Table~\ref{tab:fewshot_comparison} also isolate the
knowledge graph's contribution, because full \textsc{SHACR} and the YAML-only
Claude baseline share the identical model and differ only in whether the graph is
present. In the zero-shot setting, adding the graph lifts F1 from 0.59 to 0.79 and
cuts total classification errors (FP\,$+$\,FN) from 60 to 38, a \textbf{36.7\%}
reduction; against YAML-only Gemini (56 errors) the reduction is \textbf{32.1\%}.
Because the model is held fixed in the Claude comparison, this gap is attributable
to the graph rather than to model capability.

The error structure explains why. Without the graph, Claude reasons only over rule
text. It attains moderate precision (0.75) but poor recall (0.48), missing more
than half of all conflicts because the multi-hop dependency paths that connect
interacting rules are simply absent from the text it sees. Gemini is more balanced
(F1 0.67) but still far short of \textsc{SHACR}, and for the same reason---neither
model can follow a causal chain such as \emph{vacuum runs}~$\to$~\emph{motion
sensor fires}~$\to$~\emph{door unlocks} when no edge in the input encodes it. The
graph supplies exactly those edges: the \texttt{AFFECTS} substrate makes physical
dependencies traversable, and the hierarchical room/device/apartment structure
grounds each conflict in its correct spatial scope rather than treating rules as
free-floating text fragments. The knowledge graph, not the choice of LLM, is the
decisive architectural factor.

\subsection{Comparison with Prior Work Baselines}
\label{subsec:baseline_comparison}

The ablation isolates the graph by holding the LLM fixed; we now ask how
\textsc{SHACR} compares against established detectors from the prior literature.
We select two rule-centric baselines that are knowledge-graph--free and LLM-free
and that together bound what is achievable from rule text alone.
IoTC$^2$~\cite{al2019iotc} is the canonical formal-method detector of logical
contradictions---two rules issuing opposing actions on a shared actuator.
VISCR~\cite{nagendra2019viscr} strengthens this with a vendor-independent topology
abstraction and a mutual-exclusion filter over context-gated rules, a stronger
non-learning baseline. Both are widely cited and target the same pre-deployment
task. As neither has a public implementation, we faithfully reimplement the core
detection logic of each from its published description and run it on the identical
203-rule testbed and ground truth used throughout \S\ref{sec:eval}; this
reimplementation is the comparison's principal fairness caveat and is disclosed
explicitly.

\begin{table*}[t]
  \centering
  \caption{Conflict detection: \textsc{SHACR} vs.\ reimplemented prior-work
  baselines (rule-level, $n=203$). IoTC$^2$ and VISCR operate on rule text alone,
  without a knowledge graph or LLM, and are reimplemented from their published
  descriptions. \textsc{SHACR} (highlighted) dominates both baselines across all
  prompting configurations, even zero-shot.}
  \label{tab:baseline_comparison}
  \footnotesize
  \renewcommand{\arraystretch}{1.3}
  \begin{tabular*}{\textwidth}{@{\extracolsep{\fill}}lcccccccc@{}}
    \toprule
    \textbf{Method} & \textbf{Acc.} & \textbf{Prec.} & \textbf{Rec.} &
    \textbf{F1} & \textbf{TP} & \textbf{TN} & \textbf{FP} & \textbf{FN} \\
    \midrule
    IoTC$^2$~\cite{al2019iotc}     & 0.68 & 0.73 & 0.43 & 0.54 & 38 & 101 & 14 & 50 \\
    VISCR~\cite{nagendra2019viscr} & 0.66 & 0.70 & 0.38 & 0.49 & 33 & 101 & 14 & 55 \\
    \midrule
    \rowcolor{shragrow}
    \textbf{\textsc{SHACR} (Exp.~1, zero-shot)} &
      \textbf{0.81} & \textbf{0.78} & \textbf{0.80} & \textbf{0.79} & 70 & 95  & 20 & 18 \\
    \rowcolor{shragrow}
    \textbf{\textsc{SHACR} (Exp.~2, few-shot)} &
      \textbf{0.86} & \textbf{0.80} & \textbf{0.90} & \textbf{0.84} & 79 & 95  & 20 &  9 \\
    \rowcolor{shragrow}
    \textbf{\textsc{SHACR} (Exp.~3, few-shot)} &
      \textbf{0.96} & \textbf{1.00} & \textbf{0.90} & \textbf{0.95} & 79 & 115 &  0 &  9 \\
    \bottomrule
  \end{tabular*}
\end{table*}

Table~\ref{tab:baseline_comparison} reports the results. Even zero-shot
\textsc{SHACR} (F1 0.79) exceeds the better baseline by 0.25 F1, and balanced
few-shot \textsc{SHACR} reaches 0.95. The baselines are bounded on recall by
construction: both detect only direct action contradictions, so their ceiling is
the fraction of conflicts expressible in that form---38 of 88, or 43\%, in our
testbed. IoTC$^2$'s recall of 0.43 matches that fraction almost exactly. It
recovers essentially all logical conflicts and \emph{none} of the 50 semantic and
physical conflicts, which carry no action-level contradiction to find. VISCR's
mutual-exclusion filter additionally over-skips five genuine logical conflicts,
lowering its recall to 0.38. The lesson generalizes beyond formal methods. Every
rule-text-only approach we evaluate---the formal checkers here (F1 0.49--0.54) and
the LLMs reading raw YAML in \S\ref{subsec:ablation} (F1 0.59--0.67)---plateaus
well below \textsc{SHACR}. What breaks the ceiling is not a more capable reasoning
engine but the traversable graph substrate that exposes the multi-hop
and environment-mediated conflicts none of these methods can see.

The baselines' precision shortfall (0.73 and 0.70, versus 1.00 for \textsc{SHACR}
at Exp.~3) is equally structural. Their 14 shared false positives fall into three
benign patterns that are indistinguishable from genuine conflicts at the
rule-text level. \emph{Complementary control} pairs fire an action and its inverse
under disjoint conditions, such as irrigation that turns on when soil is dry and
off when rain is detected. \emph{Hysteresis} pairs use non-overlapping threshold
triggers that can never fire together, such as a grow-light that turns on below
200~lux and off above 800~lux. \emph{Emergency overrides} let a safety rule
deliberately countermand a routine one, such as a smoke-triggered unlock that
supersedes a routine away-lock. Telling these apart from real conflicts requires
\texttt{EVar} ranges, \texttt{AFFECTS} semantics, and context-mode exclusivity,
exactly the information the knowledge graph encodes and rule text does not, which
is why \textsc{SHACR} clears all 14. Although the perfect precision of
\S\ref{subsec:fewshot} is testbed-specific, the underlying mechanism is structural rather than tuned.

\section{Discussion}
\label{sec:disc}

\subsection{Why the Knowledge Graph Is Decisive}

The evaluation supports a single, sharp conclusion: the structured knowledge
graph---not the choice of language model or prompt---is what makes accurate
conflict detection possible. The ablation in \S\ref{subsec:ablation} holds the
model fixed and varies only the presence of the graph. The gap is large, adding the graph raises F1 from 0.59 to 0.79 and reduces total classification
errors by 36.7\% over the YAML-only baseline that uses the same model. Because
nothing but the representation changes, the improvement cannot be attributed to
model capability; it is the difference between a baseline that misses more than
half of all conflicts (recall 0.48) and a system that detects four in five
(recall 0.80), and it follows directly from making physical dependencies
traversable rather than leaving them implicit in rule text.

Compared with the graph, few-shot calibration plays a minor role, and how it acts shows why the graph is decisive. Balanced examples lift \textsc{SHACR} from F1 0.79 to
0.95 by eliminating the false positives induced by structurally similar but
benign rule pairs, yet the same examples barely move the graph-free baselines
and, for YAML-only Claude, even reduce recall. This asymmetry is not incidental.
A few-shot example can teach a decision rule, but that rule is actionable only
where its discriminating features (\texttt{AFFECTS} composition, \texttt{EVar}
ranges, and context-mode exclusivity) are present. The graph supplies these
features, so calibration takes effect. Without it, the same prompting has no
evidence to act on. Graph grounding and prompt calibration are therefore
complementary rather than interchangeable, and the latter is conditional on the
former.

\subsection{Where Detection Is Hardest, and What It Means for Deployment}

Detection is not uniform across classes, and the asymmetry is informative.
Logical conflicts are recovered most reliably because their evidence is
local---a single contradiction on a shared state node. Physical conflicts,
invisible to text-based methods, are detected with balanced precision and recall
once the multi-hop \texttt{AFFECTS} paths are explicit. Semantic conflicts remain
hardest (recall 0.52): they require the model to judge that an environmental side
effect constitutes an undesired interaction, and their misses concentrate in
apartments with long dependency chains, pointing to \texttt{EVar} modeling
completeness as the primary lever for improvement.

The practical significance is that the classes differ in consequence as well as
in difficulty. A physical conflict can simultaneously be a safety hazard, a
security hole, and a reliability failure along one \texttt{AFFECTS} path, whereas
a logical conflict typically produces only an oscillating actuator. A deployment
in a safety-critical setting---elderly or medical care, for instance---should
therefore bias the few-shot calibration toward recall on physical and
high-consequence semantic conflicts, accepting some additional false positives.
The few-shot mechanism makes this trade-off available as a lightweight, prompt-level adjustment, with no retraining and no change to the graph schema or
retrieval pipeline.

\subsection{Positioning and Future Directions}

\textsc{SHACR}'s probabilistic reasoning is deliberately aimed at the conflict
classes that formal methods cannot express. For strictly logical contradictions,
SMT solvers and finite-state verifiers offer deterministic guarantees that
\textsc{SHACR} does not; a natural hybrid would route logical conflicts to a
formal checker and reserve the GraphRAG--LLM pipeline for the semantic and
physical classes that require reasoning over natural-language descriptions and
implicit environmental dependencies---retaining formal guarantees where they are
achievable and LLM flexibility where they are not.

Three engineering directions remain. On privacy, \textsc{SHACR} already transmits
only the relevant apartment subgraph per session rather than the full graph;
locally hosted models are the principal route for deployments where data
residency is binding. On cost, the per-session LLM calls that are acceptable at
apartment scale will, at building scale, require incremental scanning that
re-analyzes only rules changed since the last cycle rather than full rescans. On
portability, production use requires platform-specific ingestion adapters to
normalize SmartThings, Apple HomeKit, Google Home, and Home Assistant formats
into the unified schema---an engineering prerequisite, not a research barrier,
but a necessary step before \textsc{SHACR} is deployed on in-the-wild rule
sets.

\section{Conclusion}
\label{sec:conclusion}
Smart home automation conflicts are fundamentally a representation problem.
Individual rules are locally correct, yet the hazards they create emerge only
through interactions across shared devices, environmental variables, and physical
topology, dependencies that remain invisible to any analysis confined to rule
text. \textsc{SHACR} overcomes this by elevating physical cause-effect
dependencies to first-class edges in a structured, unified knowledge graph.
Anchoring LLM reasoning in this relational substrate turns conflict management
from open-ended natural-language inference into deterministic multi-hop graph
traversal, executed through a safe, validated, human-in-the-loop
Scan-Explain-Repair-Validate workflow.

Our experimental evaluation confirms these design principles. On a 203-rule, 70-apartment testbed, introducing the
knowledge graph while holding the underlying LLM fixed cuts total classification
errors by 36.7\% and raises the zero-shot $F_1$-score from 0.59 to 0.79; balanced
few-shot calibration lifts it further to 0.95. The same calibration yields
negligible gains for graph-free baselines, proving that prompt engineering cannot
substitute for an explicit relational model. \textsc{SHACR} thus establishes that
structured knowledge representation, not prompt design or model choice, is the
decisive factor for dependable IoT automation, intercepting conflicts before
deployment and resolving them without burdening end users.



\vspace{-1.3cm}

\begin{IEEEbiography}{Leena A. Marghalani}
received her M.S.\
degree in Security and Information Assurance from King Fahd University of
Petroleum and Minerals (KFUPM), Dhahran, Saudi Arabia, and her B.S.\ degree in Cybersecurity and Digital Forensics from Imam
Abdulrahman Bin Faisal University (IAU), Dammam, Saudi Arabia. Her research interests
include IoT security, AI-driven cybersecurity, and usable security.
\end{IEEEbiography}

\vspace{-1.4cm}
\begin{IEEEbiography}
[{\includegraphics[width=1in,height=1.25in,clip,keepaspectratio]{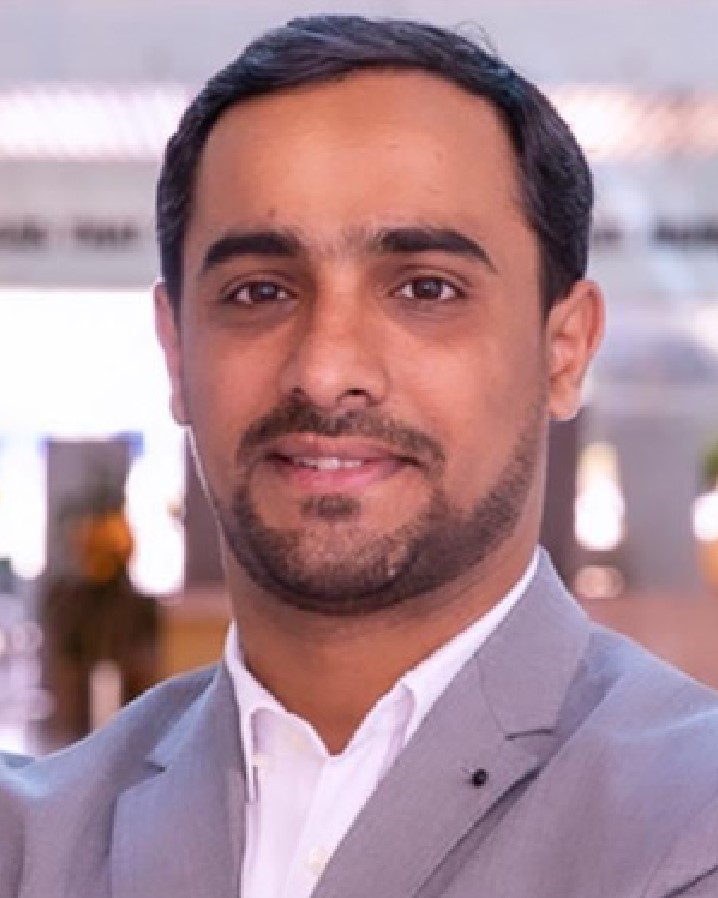}}]
{Walid Aljoby}~(Member, IEEE) received the Ph.D. degree in computer science from the National University of Singapore (NUS) in 2020, and held postdoctoral positions at NUS and Carnegie Mellon University (CMU). He is currently an Assistant Professor of computer science at the King Fahd University of Petroleum and Minerals (KFUPM). He has authored articles in the \textit{IEEE Journal on Selected Areas in Communications}, \textit{IEEE Transactions on Network and Service Management}, \textit{IEEE Open Journal of the Computer Society}, \textit{IEEE Open Journal of the Communications Society}, IEEE ICNP, IEEE NetSoft, IEEE SmartGridComm, IEEE ICC, and IEEE/IFIP NOMS. His research interests broadly span networked systems and security, with a current focus on the intersection of AI with networking and security. He is also known as Waleed Al-Gobi.
\end{IEEEbiography}
\vspace{-1.3cm}

\begin{IEEEbiography}[{\includegraphics[width=1in,height=1.25in,clip,
keepaspectratio]{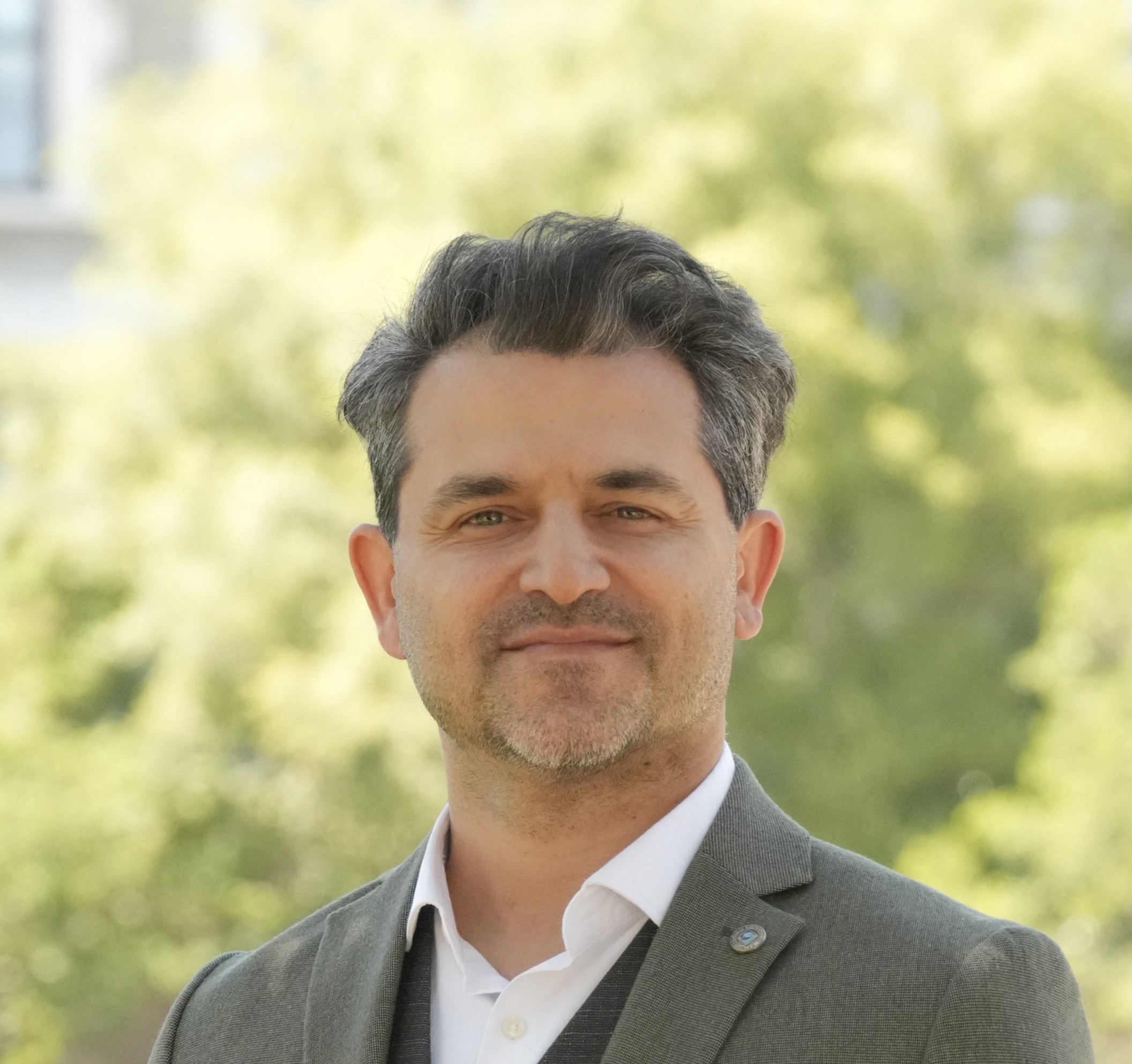}}]{Suayb S. Arslan}
(Senior Member, IEEE) received the M.Sc.\ and Ph.D.\ degrees in Electrical
and Computer Engineering from the University of California at San Diego, La
Jolla, CA, USA, in 2009 and 2012. From 2012 to 2016, he was a Senior Researcher at
Quantum Corporation, Irvine, CA, and between 2022 and 2024, he served as visiting associate Professor in the SinhaLab at MIT, Cambridge, MA, USA. He is currently a Professor in the Department of
Computer Engineering and Director of the Institute for Data Science and AI at
Bo\u{g}azi\c{c}i University, Istanbul, Türkiye. His research interests include
information theory, digital communications, cloud and quantum computing, and
AI for IoT systems. He is an Executive Editor of \textit{IoT Journal}
(Elsevier).
\end{IEEEbiography}

\vfill

\end{document}